\documentclass[12pt,a4paper]{article}
\usepackage{a4wide}
\usepackage{amsmath}
\usepackage{amssymb}
\usepackage{amsfonts}
\usepackage{epsfig}
\usepackage{subfigure}
\usepackage{exscale}
\usepackage{float}
\usepackage{bbm}
\usepackage[numbers,sort&compress]{natbib}

\newcommand{\Z}{{\mathbb{Z}}}
\newcommand{\R}{{\mathbb{R}}}
\newcommand{\C}{{\mathbb{C}}}

\newcommand{\1}{{\mathbbm{1}}}
\newcommand{\p}{\partial}

\makeatletter
\@addtoreset{equation}{section}
\makeatother

\title{Minimal Position-Velocity Uncertainty Wave Packets
in Relativistic and Non-relativistic Quantum Mechanics}

\author{M.\ H.\ Al-Hashimi and U.-J.\ Wiese \\ \\
Albert Einstein Center for Fundamental Physics \\
Institute for Theoretical Physics, Bern University \\
Sidlerstrasse 5, CH-3012 Bern, Switzerland \\ \\}

\begin{document} 

\maketitle

\vspace{-1cm}

\begin{abstract} \normalsize

We consider wave packets of free particles with a general energy-momentum 
dispersion relation $E(p)$. The spreading of the wave packet is determined by 
the velocity $v = \p_p E$. The position-velocity uncertainty relation 
$\Delta x \Delta v \geq \frac{1}{2} |\langle \p_p^2 E \rangle|$ is saturated by
minimal uncertainty wave packets $\Phi(p) = A \exp(- \alpha E(p) + \beta p)$.
In addition to the standard minimal Gaussian wave packets corresponding to the
non-relativistic dispersion relation $E(p) = p^2/2m$, analytic calculations are
presented for the spreading of wave packets with minimal position-velocity 
uncertainty product for the lattice dispersion relation 
$E(p) = - \cos(p a)/m a^2$ as well as for the relativistic dispersion relation 
$E(p) = \sqrt{p^2 + m^2}$. The boost properties of moving relativistic wave 
packets as well as the propagation of wave packets in an expanding Universe are
also discussed.

\end{abstract}

\newpage
 
\section{Introduction}

The spreading of Gaussian wave packets is a standard topic discussed in almost
any textbook on quantum mechanics. While there still are interesting 
investigations and applications of wave packet dynamics \cite{Gar95,And08}, it 
may seem unlikely that anything new can be said theoretically about a topic as 
elementary as this. Indeed, some of the results that will be presented below 
have been discussed before. Still, to the best of our knowledge, most of our 
results, although easy to derive, are new and seem to have escaped the 
attention of quantum physicists. 
The main goal of this paper is to present a general discussion of wave packet 
spreading, which generalizes the standard Gaussian wave packet describing a 
free non-relativistic particle to other minimal position-velocity uncertainty 
wave packets with a general energy-momentum dispersion relation $E(p)$. 

As was first discussed by Ehrenfest \cite{Ehr27}, a free non-relativistic 
particle moving in one spatial dimension shows the following time-dependence of
the position expectation value
\begin{equation}
\langle x \rangle(t) = \langle x \rangle(0) + \langle v \rangle t,
\end{equation}
where $v = \p_p E$ is the corresponding velocity. It is easy to show that this
relation holds for any (relativistic or non-relativistic) dispersion relation 
$E(p)$. Similarly, one can show that the position uncertainty 
$\Delta x(t) = \sqrt{\langle x^2 \rangle(t) - \langle x \rangle(t)^2}$ varies as
\begin{equation}
\Delta x(t)^2 = \Delta x(0)^2 + \left[\langle v x + x v \rangle(0) -
2 \langle v \rangle \langle x \rangle(0)\right] t + (\Delta v)^2 \ t^2.
\end{equation}
This implies that the asymptotic speed of spreading for general wave packets is
given by $\Delta v = \sqrt{\langle v^2 \rangle - \langle v \rangle^2}$. 
Furthermore, if position and velocity are correlated such that initially 
$\langle v x + x v \rangle(0) < 2 \langle v \rangle \langle x \rangle(0)$, 
the wave packet is shrinking before it begins to spread.

It is interesting to investigate wave packets of minimal position-velocity
uncertainty product $\Delta x \Delta v$. Putting $\hbar = 1$, one easily 
derives the generalized position-velocity uncertainty relation
\begin{equation}
\label{uncertainty}
\Delta x \Delta v \leq \frac{1}{2} \left\vert 
\left\langle \p_p^2 E \right\rangle \right\vert,
\end{equation}
which is valid for any dispersion relation, and which reduces to the standard
Heisenberg uncertainty relation $\Delta x \Delta v \leq \frac{1}{2m}$ for a
non-relativistic particle with energy $E(p) = p^2/2m$. For a particle hopping
between neighboring sites on a lattice with spacing $a$, the dispersion relation
is $E(p) = - \cos(pa)/m a^2$ and the uncertainty relation then takes the form
\begin{equation}
\Delta x \Delta v \leq \frac{a^2}{2} \vert \langle E \rangle \vert.
\end{equation}
On the other hand, (also putting $c = 1$) for a relativistic particle with 
$E(p) = \sqrt{p^2 + m^2}$ the uncertainty relation reduces to
\begin{equation}
\Delta x \Delta v \leq \frac{m^2}{2} \langle E^{-3} \rangle.
\end{equation}
For a general dispersion relation, minimal uncertainty wave packets $\Phi(p)$
saturate the inequality eq.(\ref{uncertainty}) and obey the equation
\begin{equation}
\left(\p_p + \alpha v - \beta\right) \Phi(p) = 0,
\end{equation}
with 
\begin{equation}
\alpha = \frac{1}{2 (\Delta v)^2} \left\vert \left\langle \p_p^2 E 
\right\rangle \right\vert,
\quad \beta = \alpha \langle v \rangle - i \langle x \rangle.
\end{equation}
In momentum space, they take the form
\begin{equation}
\Phi(p) = A \exp(- \alpha E(p) + \beta p).
\end{equation}
In this paper,  minimal uncertainty wave packets are constructed explicitly and
their time-evolution is calculated analytically, both for the lattice and for 
the relativistic dispersion relation.

Several results of this paper belong to relativistic quantum mechanics. It 
should be noted that both the Dirac and the Klein-Gordon equation belong to
quantum field theory and not to relativistic quantum mechanics. In particular, 
it is well-known that, due to the existence of negative energy solutions, a 
purely quantum mechanical single particle interpretation of these equations 
leads to problems such as the Klein paradox. When we discuss the relativistic 
quantum mechanics of a single free particle, we consider the Hamiltonian
\begin{equation}
\label{relH}
H = \sqrt{p^2 + m^2},
\end{equation}
which only has positive energy solutions. While we know that the correct
description of Nature at the most fundamental level accessible today is provided
by the standard model of particle physics --- which is a relativistic quantum 
field theory --- there is nothing wrong with studying the Hamiltonian of 
eq.(\ref{relH}). In particular, while the corresponding square root would 
conflict with locality in quantum field theory, it is perfectly acceptable in 
the framework of single particle quantum mechanics. In fact, as will be 
discussed in appendix B, the above Hamiltonian correctly describes the single 
particle states of a free scalar quantum field theory. As borne out by the 
Reeh-Schlieder theorem \cite{Ree61}, the localization of particles is a subtle 
issue in quantum field theory. The interpretation of the results obtained in 
relativistic quantum mechanics must take this into account. In particular, 
apparent violations of Einstein causality in relativistic quantum mechanics 
turn out to be unproblematical when viewed from this perspective
\cite{Fle65,Heg74,Heg80,Heg85,Ros87,Mos90,Heg01,Bar03,Bus05,Bus06}.

In contrast to quantum field theory, relativistic quantum mechanics (as
characterized above) seems not to be a well-studied subject. This is a pity, 
because relativistic quantum mechanics may help to bridge the large gap between 
non-relativistic quantum mechanics and relativistic quantum field theory that 
makes learning the latter rather non-trivial. Also for pedagogical reasons, it 
thus seems worthwhile to study relativistic quantum mechanics --- even of just a
single particle. Due to Lorentz invariance, the possible interactions in 
relativistic quantum mechanics are highly constrained. In particular, if one 
attempts to include interactions only in the Hamiltonian and in the boost 
operator (but not in the momentum or angular momentum operators) of a 
multi-particle system, at the classical level Leutwyler has proved that only 
the free theory is consistent with Lorentz invariance \cite{Leu65}. The same 
was first shown for two particles by Currie, Jordan, and Sudarshan 
\cite{Cur63}. By including the interaction in the 
momentum (and not in the boost operator), interesting relativistic systems with
a fixed number of interacting particles have been constructed and investigated 
in detail \cite{Rui80,Rui86,Rui01}. Here we concentrate on wave packet spreading
of a single free relativistic particle. Although in this paper we do not focus 
on applications, the resulting expressions may be useful in studies of neutrino
wave packets \cite{Giu91,Giu04}. For this reason, we also consider the 
spreading of wave packets in an expanding Universe.

The rest of the paper is organized as follows. In section 2 the time-dependence
of position expectation values is considered for a general dispersion relation,
the corresponding position-velocity uncertainty relation is derived, and the
general minimal uncertainty wave packet is constructed. Section 3 investigates 
wave packet spreading in the non-relativistic case, both in the continuum and
for a lattice dispersion relation. The spreading of minimal position-velocity 
uncertainty wave packets for a free relativistic particle is discussed in 
detail in section 4. Section 5 extends the discussion to wave packets 
propagating in an expanding Universe. Finally, section 6 contains our 
conclusions. Some non-trivial expectation values are evaluated in appendix A. 
The relation of relativistic quantum mechanics to quantum field theory and the 
related issues of particle localization are reviewed in appendix B.

\section{Spreading of General Wave Packets}

In this section we investigate general properties of spreading wave packets.
We also derive a position-velocity uncertainty relation and we consider wave
packets with a minimal position-velocity uncertainty product. While the results 
presented in this section are quite elementary, except for those in subsection 
2.1, we have not been able to locate them in the physics literature. Although 
it seems likely that some of the material has been discussed elsewhere, it 
seems not to be well-known and may thus be worth studying in some detail.

\subsection{Time-Evolution of Position Expectation Values}

Let us consider a single free particle in one spatial dimension with momentum 
$p$ and energy $E(p)$. The time-evolution of an initial wave function $\Psi(p)$
in momentum space is then given by
\begin{equation}
\Psi(p,t) = \Psi(p) \exp(- i E(p) t).
\end{equation}
The corresponding wave function in coordinate space takes the form
\begin{eqnarray}
\Psi(x,t)&=&\frac{1}{2 \pi} \int dp \ \Psi(p,t) \exp(i p x) =
\frac{1}{2 \pi} \int dp \ \Psi(p) \exp(- i E(p) t + i p x) \nonumber \\
&=&\int dx' \ G(x - x',t) \ \Psi(x',0),
\end{eqnarray}
with the Green's function given by
\begin{equation}
\label{Green}
G(x,t) = \frac{1}{2 \pi} \int dp \ \exp(- i E(p) t + i p x).
\end{equation}
Using $x = i \p_p$ one then obtains
\begin{eqnarray}
\label{averagex}
\langle x \rangle(t)&=&\frac{1}{2 \pi} \int dp \ \Psi(p,t)^* i \p_p \Psi(p,t) =
\frac{1}{2 \pi} \int dp \ \Psi(p)^*\left(i \p_p + \p_p E \ t \right)\Psi(p) 
\nonumber \\
&=&\langle x \rangle(0) + \langle v \rangle t,
\end{eqnarray}
where $v = \p_p E$ is the particle's velocity. Similarly, one finds
\begin{eqnarray}
\label{averagex2}
\langle x^2 \rangle(t)&=&\frac{1}{2 \pi} \int dp \ \Psi(p,t)^* (i \p_p)^2 
\Psi(p,t) \nonumber \\
&=&\frac{1}{2 \pi} \int dp \ \Psi(p)^*\left[(i \p_p)^2 + 
\left(2 \p_p E \ i \p_p + i \p_p^2 E \right) t + 
\left(\p_p E \right)^2 t^2 \right] \Psi(p) 
\nonumber \\
&=&\langle x^2 \rangle(0) + \langle v x + xv \rangle(0) t + 
\langle v^2 \rangle t^2,
\end{eqnarray}
where we have used 
\begin{equation}
[x,v] = \left[i \p_p,\p_p E \right] = i \p_p^2 E.
\end{equation}
Combining eq.(\ref{averagex}) with eq.(\ref{averagex2}) one then obtains
\begin{equation}
\label{spreading}
\Delta x(t)^2 = \Delta x(0)^2 + \left[\langle v x + x v \rangle(0) -
2 \langle v \rangle \langle x \rangle(0)\right] t + (\Delta v)^2 \ t^2.
\end{equation}
The sign of the connected position-velocity correlation 
$\langle v x + x v \rangle(0) - 2 \langle v \rangle \langle x \rangle(0)$ 
determines whether the wave packet is initially spreading or shrinking. 
Asymptotically, for large times the packet is spreading with the velocity 
$\Delta v$, i.e.\ the velocity uncertainty determines the velocity of 
spreading. The time-dependence of moments of the position operator is 
well-known \cite{Ehr27,Gar95} and has also been considered, for example, in 
\cite{Bai72}.

\subsection{General Position-Velocity Uncertainty Relation}

As we have just seen, the spreading of general wave packets is controlled by the
uncertainties $\Delta x$ and $\Delta v$ of position and velocity. This suggests
to consider packets with a minimal position-velocity uncertainty product.
Before we construct such wave packets, let us derive a generalization of the
non-relativistic Heisenberg uncertainty relation
\begin{equation}
\label{Heis}
\Delta x \Delta v = \frac{1}{m} \Delta x \Delta p \geq \frac{1}{2m} 
\end{equation}
to an arbitrary dispersion relation $E(p)$ with velocity $v = \p_p E$. For this 
purpose, we define the operator 
\begin{equation}
a = - i x + \alpha v - \beta = \p_p + \alpha v - \beta, 
\end{equation}
(with $\alpha \in \R$ and $\beta = \beta_r + i \beta_i \in \C$ as arbitrary
adjustable parameters) and we evaluate
\begin{eqnarray}
\label{ineq}
\langle a^\dagger a \rangle&=&\langle\left(i x + \alpha v - \beta^*\right)
\left(- i x + \alpha v - \beta\right)\rangle \nonumber \\
&=&\langle x^2 + \alpha^2 v^2 + |\beta|^2 + i \alpha [x,v] -
i (\beta - \beta^*) x - \alpha (\beta + \beta^*) v \rangle \nonumber \\
&=&\langle x^2 \rangle + \alpha^2 \langle v^2 \rangle + \beta_r^2 + \beta_i^2 -
\alpha \left\langle \p_p^2 E \right\rangle + 
2 \beta_i \langle x \rangle - 2 \alpha \beta_r \langle v \rangle \geq 0.
\end{eqnarray}
By construction $\langle a^\dagger a \rangle \geq 0$. In order to obtain the 
most stringent bound, we now vary the free parameters $\alpha$, $\beta_r$, and
$\beta_i$ such that $\langle a^\dagger a \rangle$ is minimized. This implies the
conditions
\begin{eqnarray}
&&\frac{\p \langle a^\dagger a \rangle}{\p \alpha} = 
2 \alpha \langle v^2 \rangle - \left\langle \p_p^2 E \right\rangle - 
2 \beta_r \langle v \rangle = 0, \nonumber \\
&&\frac{\p \langle a^\dagger a \rangle}{\p \beta_r} = 2 \beta_r - 2 \alpha
\langle v \rangle = 0, \nonumber \\
&&\frac{\p \langle a^\dagger a \rangle}{\p \beta_i} = 2 \beta_i + 
2 \langle x \rangle = 0,
\end{eqnarray}
which are satisfied when
\begin{equation}
\label{parameters}
\alpha = \frac{1}{2 (\Delta v)^2} \left\langle \p_p^2 E \right\rangle,
\quad \beta_r = \alpha \langle v \rangle, \quad \beta_i = - \langle x \rangle.
\end{equation}
Inserting these values in eq.(\ref{ineq}) one obtains
\begin{equation}
\label{Heisg}
(\Delta x)^2 - \frac{1}{4 (\Delta v)^2} \left\langle \p_p^2 E 
\right\rangle^2 \geq 0 \ \Rightarrow \ \Delta x \Delta v \geq 
\frac{1}{2} \left\vert \left\langle \p_p^2 E \right\rangle \right\vert.
\end{equation}
Indeed, in the non-relativistic case, $E(p) = p^2/2m$, this reduces to the
standard Heisenberg uncertainty relation eq.(\ref{Heis}). Obviously, this is
just a special case of the general uncertainty relation
$\Delta A \Delta B \leq \frac{1}{2} \vert\langle [A,B] \rangle\vert$.

\subsection{Minimal Position-Velocity Uncertainty Wave Packets}

By construction, it is clear that wave packets $\Phi(p)$ with a minimal 
position-velocity uncertainty product $\Delta x \Delta v$, which saturate the 
inequality eq.(\ref{Heisg}), must satisfy 
\begin{equation}
a \Phi(p) = \left(\p_p + \alpha v - \beta\right) \Phi(p) = 0. 
\end{equation}
This equation is easy to solve and one obtains
\begin{equation}
\label{general}
\Phi(p) = A \exp(- \alpha E(p) + \beta p).
\end{equation}
Using eq.(\ref{parameters}) and assuming that $\langle \p_p^2 E \rangle > 0$,
this can also be expressed as
\begin{equation}
\Phi(p) = A \exp\left(- \frac{\Delta x}{\Delta v}
\left[E(p) - p \langle v \rangle\right] - i p \langle x \rangle\right).
\end{equation}
In coordinate space, a minimal uncertainty wave packet takes the form
\begin{eqnarray}
\Phi(x,t)&=&\frac{1}{2 \pi} \int dp \ \Phi(p) 
\exp\left[- i E(p) t + i p x\right] \nonumber \\
&=&\frac{A}{2 \pi} \int dp \ 
\exp\left[- i E(p) (t - i \alpha) + i p (x - i \beta)\right] \nonumber \\
&=&A \ G(x - i \beta,t - i \alpha),
\end{eqnarray}
and is thus given by analytic continuation of the Green's function $G(x,t)$ of 
eq.(\ref{Green}). It is straightforward to show that for a wave packet with 
minimal position-velocity uncertainty the initial position-velocity correlation
vanishes, i.e.\ 
$\langle v x + x v \rangle(0) = 2 \langle v \rangle \langle x \rangle(0)$. 
The general formula eq.(\ref{spreading}) describing wave packet spreading then 
reduces to
\begin{equation}
\Delta x(t)^2 = \Delta x(0)^2 + (\Delta v)^2 \ t^2.
\end{equation}
Since for a free particle $\langle v \rangle$ and $\langle v^2 \rangle$ are
time-independent, a wave packet that initially has a minimal position-velocity 
uncertainty will obviously not maintain a minimal uncertainty product as time 
evolves. In fact, one obtains
\begin{equation}
\Delta x(t) \Delta v = 
\sqrt{\frac{1}{4} \left\langle \p_p^2 E \right\rangle^2 + 
(\Delta v)^4 \ t^2}.
\end{equation}

\subsection{Generalization to Higher Dimensions}

It is straightforward to extend the results of the previous subsections to
higher dimensions. Let us consider a free particle with momentum $\vec p$ and
energy $E(\vec p)$ moving in $d$ dimensions. The time-evolution of an initial
momentum space wave function $\Psi(\vec p)$ is then given by
\begin{equation}
\Psi(\vec p,t) = \Psi(\vec p) \exp(- i E(\vec p) t),
\end{equation}
and the corresponding coordinate space wave function takes the form
\begin{equation}
\Psi(\vec x,t) = \frac{1}{(2 \pi)^d} \int d^dp \ \Psi(\vec p)
\exp(- i E(\vec p) t + i \vec p \cdot \vec x) = \int d^dx' \ 
G(\vec x - {\vec x}',t) \Psi({\vec x}',0),
\end{equation}
where the Green's function is given by
\begin{equation}
G(\vec x,t) = \frac{1}{(2 \pi)^d} \int d^dp \
\exp(- i E(\vec p) t + i \vec p \cdot \vec x). 
\end{equation}
In $d$ dimensions we have
\begin{equation}
\vec v = \vec \nabla_p E, \quad [x_i,v_j] = 
\left[i \p_{p_i},\p_{p_j} E \right] = i \p_{p_i} \p_{p_j} E,
\end{equation}
where $\vec \nabla_p$ denotes the gradient in momentum space, and one then 
obtains
\begin{eqnarray}
&&\langle \vec x \rangle(t) = \langle \vec x \rangle(0) + 
\langle \vec v \rangle t,
\nonumber \\
&&\langle {\vec x}^{\! \ 2} \rangle(t) = \langle {\vec x}^{\! \ 2} \rangle(0) + 
\langle \vec v \cdot \vec x + \vec x \cdot \vec v \rangle(0) t +
\langle {\vec v}^{\! \ 2} \rangle t^2,
\nonumber \\
&&\Delta x(t)^2 = \Delta x(0)^2 + 
\left[\langle \vec v \cdot \vec x + \vec x \cdot \vec v \rangle(0) -
2 \langle \vec v \rangle \cdot \langle \vec x \rangle(0)\right] t + 
(\Delta v)^2 t^2,
\end{eqnarray}
with 
$\Delta x = \sqrt{\langle {\vec x}^{\! \ 2} \rangle - \langle \vec x \rangle^2}$
and $\Delta v = \sqrt{\langle {\vec v}^{\! \ 2} \rangle - 
\langle \vec v \rangle^2}$.

In order to derive the generalized position-velocity uncertainty relation we
define
\begin{equation}
\vec a = - i \vec x + \alpha \vec v - \vec \beta = 
\vec \nabla_p + \alpha \vec v - \vec \beta,
\end{equation}
Minimizing $\langle {\vec a}^{\ \dagger} \cdot \vec a \rangle$ one obtains
\begin{equation}
\alpha = \frac{1}{2 (\Delta v)^2} \left\langle \Delta_p E \right\rangle,
\quad \vec \beta = \alpha \langle \vec v \rangle - i \langle \vec x \rangle,
\end{equation}
which then implies
\begin{equation}
(\Delta x)^2 (\Delta v)^2 \geq 
\frac{1}{4} \left\langle \Delta_p E \right\rangle^2,
\end{equation}
where $\Delta_p$ denotes the Laplace operator in momentum space. For the 
non-relativistic dispersion relation $E(\vec p) = {\vec p}^{\! \ 2}/2m$ one 
obtains the $d$-dimensional Heisenberg uncertainty relation
\begin{equation}
\Delta x \Delta v \geq \frac{d}{2 m},
\end{equation}
while for the relativistic dispersion relation $E(p) = 
\sqrt{{\vec p}^{\! \ 2} + m^2}$ one obtains
\begin{equation}
\Delta x \Delta v \geq \frac{d m^2}{2} \left\langle E^{-3} \right\rangle.
\end{equation}

A minimal position-velocity uncertainty wave packet must obey
\begin{equation}
\vec a \Phi(\vec p) = \left(\vec \nabla_p + \alpha \vec v - \vec \beta\right) 
\Phi(\vec p) = 0, 
\end{equation}
which is solved by
\begin{equation}
\Phi(\vec p) = A \exp(- \alpha E(\vec p) + \vec \beta \cdot \vec p).
\end{equation}
In coordinate space, a minimal uncertainty wave packet then takes the form
\begin{eqnarray}
\Phi(\vec x,t)&=&\frac{1}{(2 \pi)^d} \int d^dp \ \Phi(\vec p) 
\exp\left[- i E(\vec p) t + i \vec p \cdot \vec x\right] \nonumber \\
&=&\frac{A}{(2 \pi)^d} \int d^dp \ 
\exp\left[- i E(\vec p) (t - i \alpha) + i \vec p \cdot (\vec x - i \vec \beta)
\right] \nonumber \\
&=&A \ G(\vec x - i \vec \beta,t - i \alpha),
\end{eqnarray}

\section{Spreading of Non-relativistic Wave Packets in the Continuum and on a Lattice}

In this section we consider free non-relativistic particles either in the
continuum, i.e.\ with $E(p) = p^2/2m$, or on a lattice with spacing $a$ and
$E(p) = - \cos(p a)/m a^2$.

\subsection{Spreading of Standard Gaussian Wave Packets}

Although this subsection contains completely standard textbook material, we 
like to include it, in order to ease the transition to the relativistic case
discussed in the next section. Let us consider a non-relativistic free
particle with $E(p) = p^2/2m$. The minimal uncertainty wave packet then takes
the standard Gaussian form
\begin{equation}
\Phi(p) = A \exp\left(- \alpha \frac{p^2}{2m} + \beta p\right).
\end{equation}
For $\beta_i = 0$ the corresponding expectation values are given by
\begin{eqnarray}
&&\langle x \rangle = 0, \quad \langle x^2 \rangle = (\Delta x)^2 =
\frac{\alpha}{2 m}, \nonumber \\
&&\langle v \rangle = \frac{\beta}{\alpha}, \quad 
\langle v^2 \rangle = \frac{1}{2 m \alpha} + \frac{\beta^2}{\alpha^2}, \quad 
(\Delta v)^2 = \frac{1}{2 m \alpha}.
\end{eqnarray}
For $\beta_i \neq 0$ the wave packet is just shifted in space by $- \beta_i$. 
The wave function $\Phi(p)$ translates into the coordinate space form 
$\Phi(x,t) = A G(x - i \beta,t - i \alpha)$ with the Green's function given by
\begin{equation}
G(x,t) = \sqrt{\frac{m}{2 \pi i t}} \exp\left(\frac{i m x^2}{2 t}\right).
\end{equation}
The spreading of two minimal position-velocity uncertainty wave packets is
illustrated in figure 1.
\begin{figure}[htb]
\begin{center}
\includegraphics[width=0.3\textwidth,angle=-90]{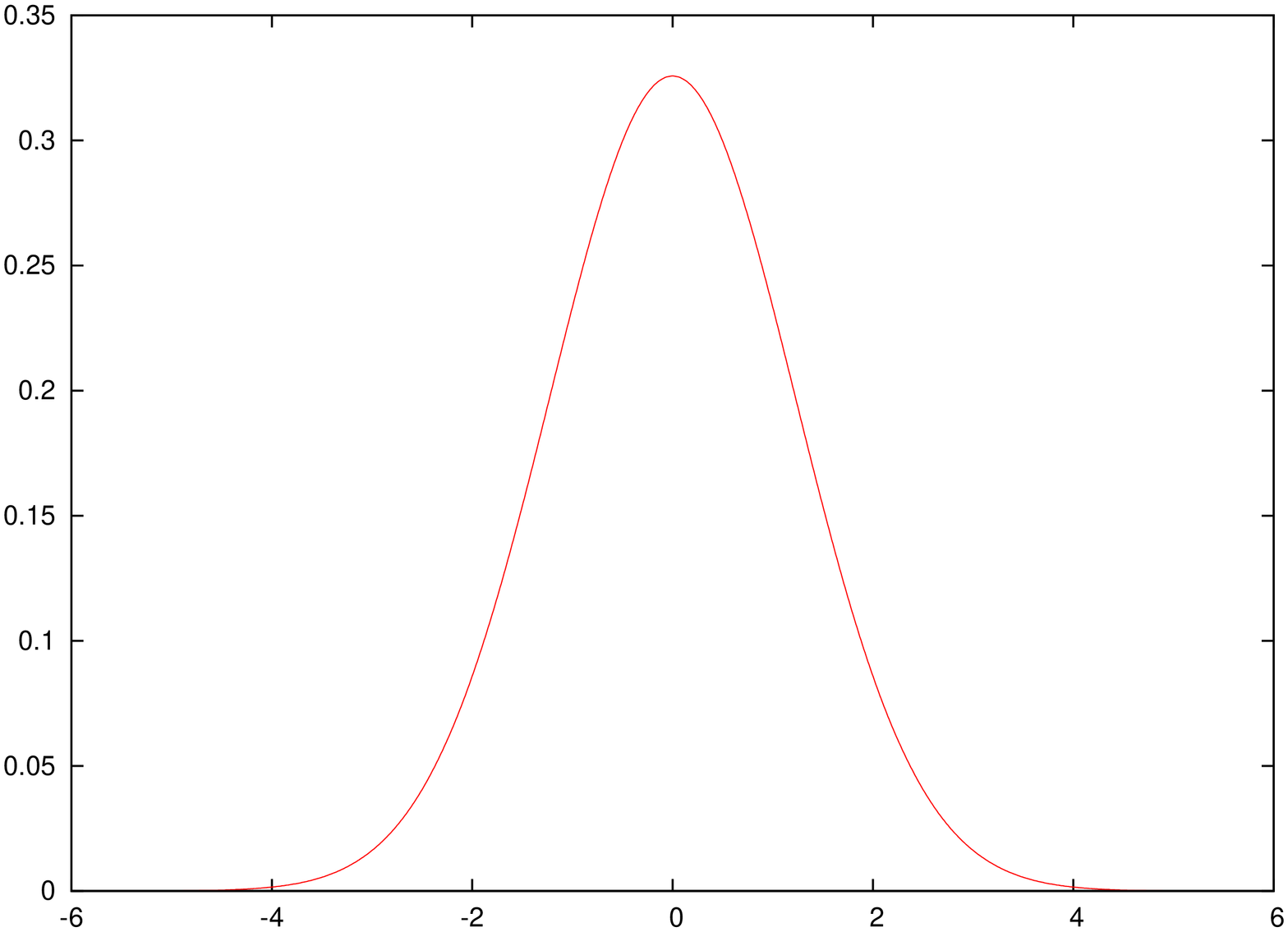} \hskip-0.5cm
\includegraphics[width=0.41\textwidth,angle=-90]{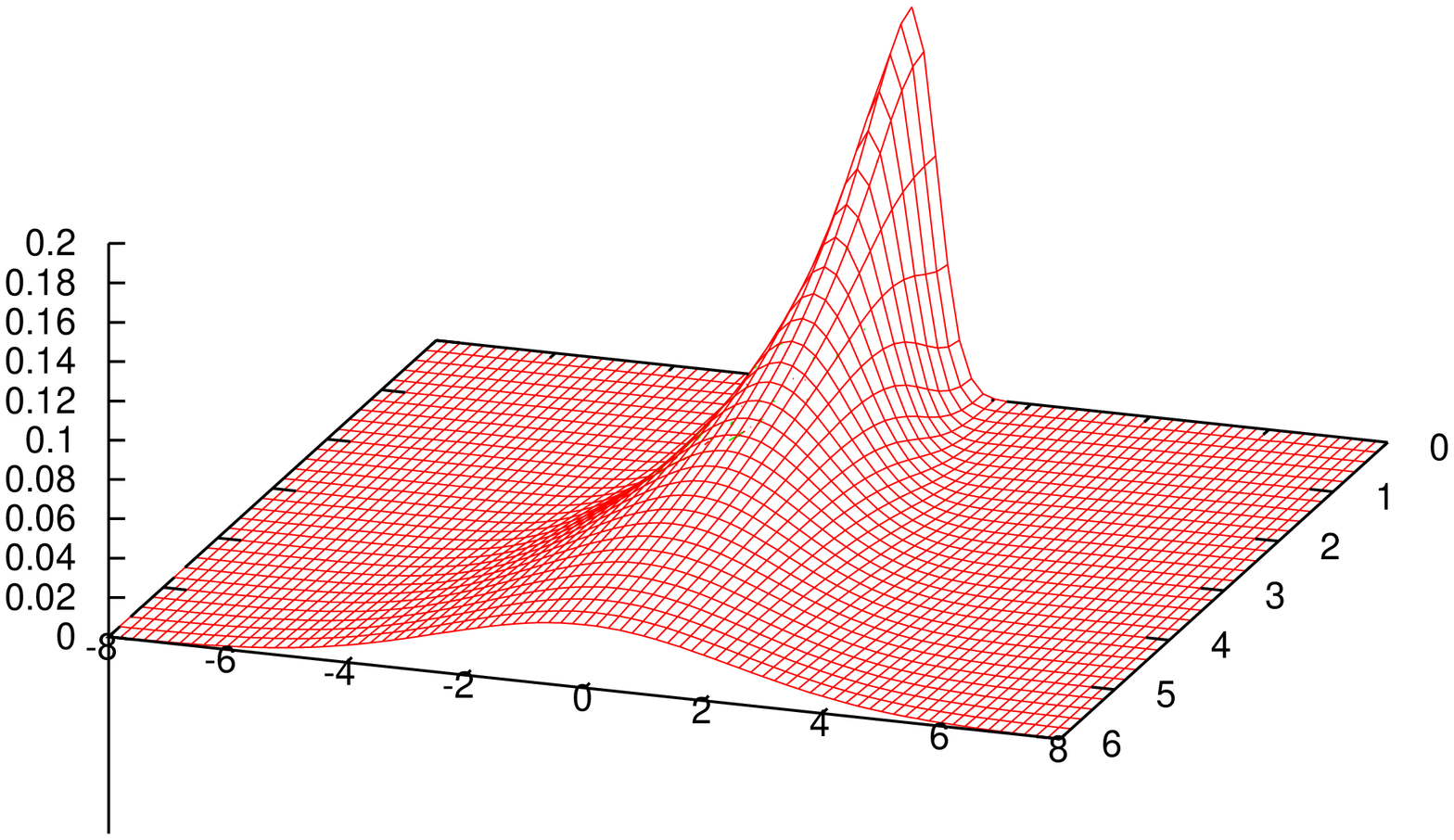}
\includegraphics[width=0.3\textwidth,angle=-90]{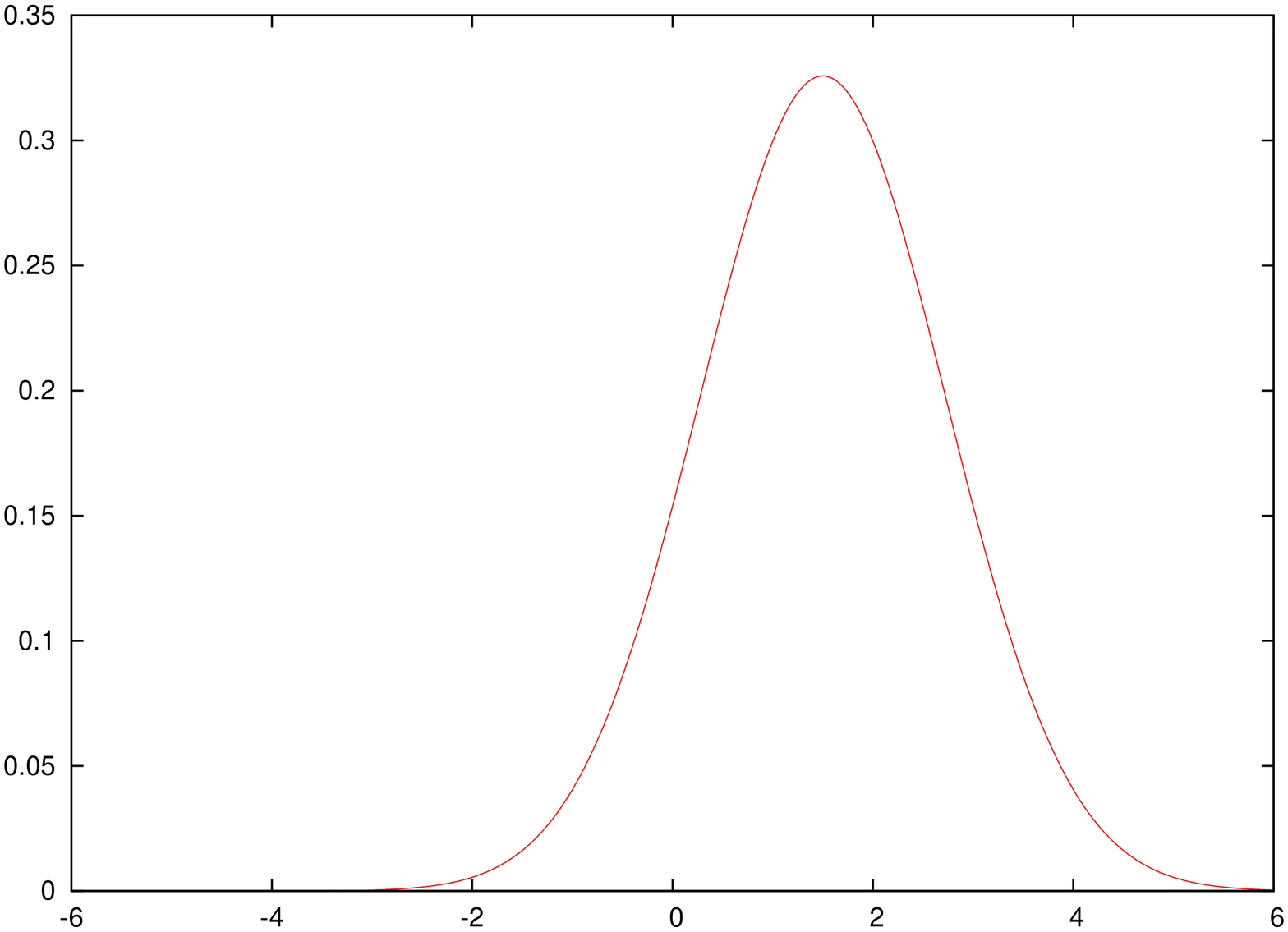} \hskip-0.5cm
\includegraphics[width=0.41\textwidth,angle=-90]{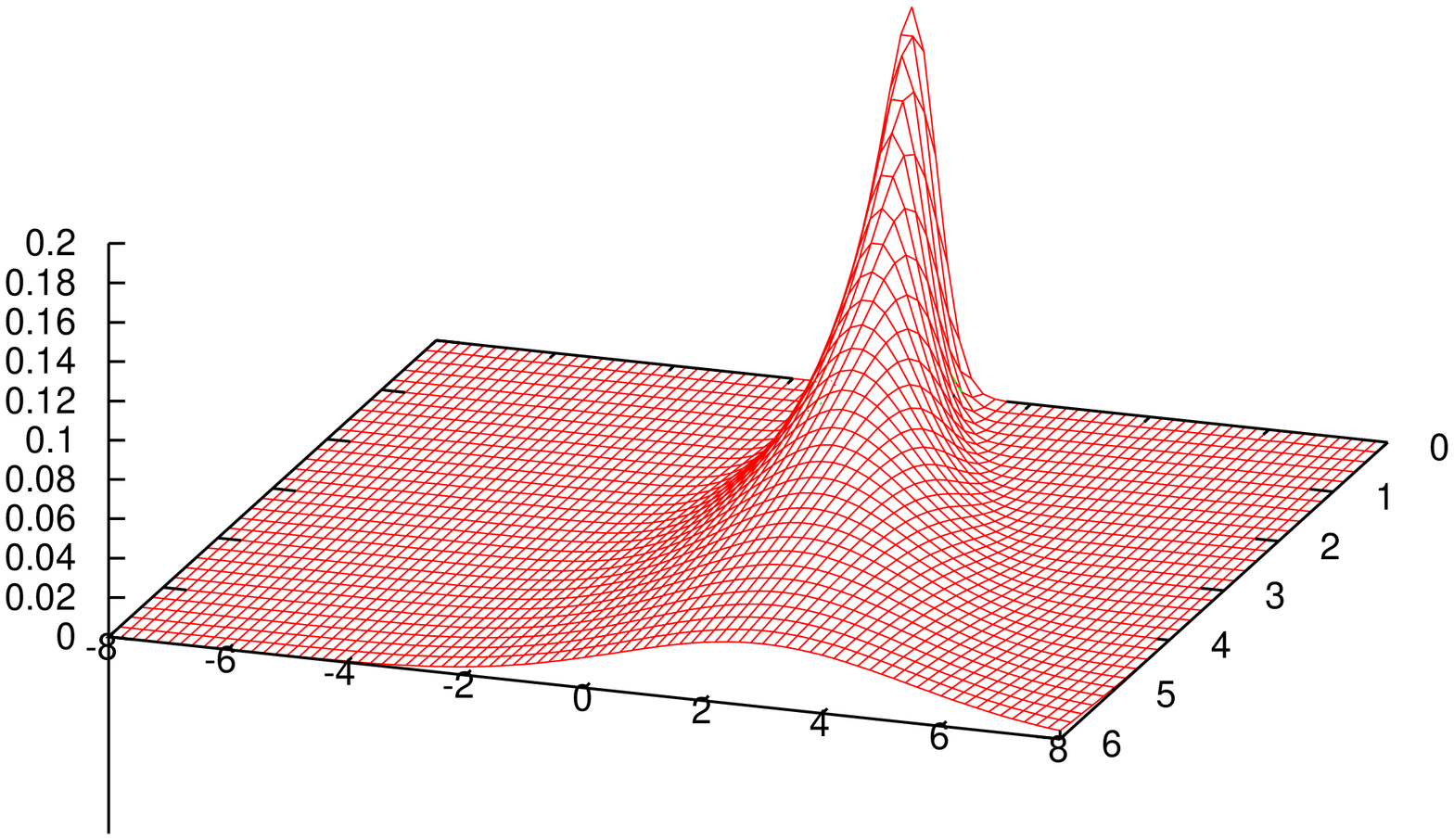}
\vskip-1cm
\end{center}\vskip-1cm
\caption{\it Probability distributions of two non-relativistic minimal 
position-velocity uncertainty wave packets in momentum space (left) and 
spreading in coordinate space as a function of time (right) for $\alpha = 1$, 
$\beta = 0$ (top) and for $\alpha = 1$, $\beta = 1/2$ (bottom) with $m=3$.}
\end{figure}

As a preparation for the relativistic case to be discussed in section 4,
let us consider the Galilean boost properties of the spreading Gaussian wave 
packet. The generators of the Galilean group are the Hamiltonian $H$, the 
momentum $P$, and the Galilean boost $M$, which are given by
\begin{equation}
H = \frac{p^2}{2m}, \ P = p, \ M = m x,
\end{equation}
and which obey the commutation relations
\begin{equation}
[H,P] = 0, \ [M,P] = i m, \ [M,H] = i P.  
\end{equation}
The unitary transformation
\begin{equation}
U(u) = \exp(- i u M) = \exp(- i u m x) = \exp(u m \p_p),
\end{equation}
which implements a boost to an inertial frame moving with the velocity $u$, acts
as a shift-operator on an arbitrary momentum space wave function, i.e.\
\begin{equation}
\Psi_b(p') = U(u) \Psi(p') = \Psi(p), \quad p' = p - u m.
\end{equation}
In particular, for a minimal uncertainty wave packet we obtain
\begin{eqnarray}
\Phi_b(p')&=&U(u) \Phi(p') = \Phi(p) = 
A \exp\left(- \alpha \frac{(p' + u m)^2}{2m} + \beta (p' + u m)\right) 
\nonumber \\
&=&A' \exp\left(- \alpha' \frac{{p'}^{\! \ 2}}{2m} + \beta' p'\right),
\end{eqnarray}
with the parameters after the boost given by
\begin{equation}
\alpha' = \alpha, \ \beta' = \beta - u \alpha \ \Rightarrow \
 \Delta v' = \Delta v, \ \langle v' \rangle = \langle v \rangle - u.
\end{equation}
Both  $\Delta x$ and $\Delta v$ remain unchanged after the boost, and hence a
minimal position-velocity uncertainty wave packet has minimal uncertainty also
from the point of view of a moving observer. As we will see later, this is 
different in the relativistic case.

\subsection{Spreading of Wave Packets on a Lattice}

Let us now consider a non-relativistic particle hopping between neighboring
sites on a lattice with spacing $a$. The energy-momentum dispersion relation is
then given by
\begin{equation}
E(p) = - \frac{\cos(p a)}{m a^2}.
\end{equation}
In this case, $\p_p^2 E = - a^2 E(p)$ such that the position velocity 
uncertainty relation takes the form
\begin{equation}
\Delta x \Delta v \leq \frac{a^2}{2} \vert \langle E \rangle \vert.
\end{equation}
The general solution of the minimal position-velocity uncertainty wave packet
given by $\Phi(p) = A \exp(- \alpha E(p) + \beta p)$ is not periodic over the 
Brillouin zone $]-\pi/a,\pi/a]$, and is thus not appropriate for the particle 
hopping on the lattice. Indeed, the periodicity requirement $\Phi(p + 2\pi/a) = 
\Phi(p)$ implies $\beta_r = 0$, and $\beta_i/a \in \Z$. Consequently, minimal 
position-velocity uncertainty wave packets on the lattice must obey
$\langle v \rangle = 0$ as well as $\langle x \rangle/a \in \Z$, i.e.\ they do
not move sideways and are centered at a lattice point. In contrast to the 
particle moving in the continuum, a moving wave packet on the lattice cannot 
have a minimal uncertainty product. This is due to the absence of Galilean 
symmetry, which is explicitly broken by the lattice. In fact, the lattice 
defines a preferred reference frame, which is the one in which minimal 
uncertainty wave packets (as well as the lattice itself) are at rest. The 
expectation values of various operators for these wave packets are worked out 
in appendix A and (for $\beta_i = 0$) one obtains
\begin{eqnarray}
&&\langle x \rangle = 0, \quad 
\langle x^2 \rangle = (\Delta x)^2 = \frac{\alpha I_1}{2 m I_0}, \nonumber \\
&&\langle v \rangle = 0, \quad 
\langle v^2 \rangle = (\Delta v)^2 = \frac{I_1}{2 m \alpha I_0}, \nonumber \\
&&\langle E \rangle = - \frac{I_1}{m a^2 I_0}, \quad 
\langle E^2 \rangle = \frac{1}{m^2 a^4}
\left(1 - \frac{m a^2 I_1}{2 \alpha I_0}\right), \nonumber \\
&&(\Delta E)^2 = \frac{1}{m^2 a^4}
\left(1 - \frac{m a^2 I_1}{2 \alpha I_0} - \frac{I_1^2}{I_0^2}\right). 
\end{eqnarray}
Here $I_0$ and $I_1$ are modified Bessel functions of degree zero and one with
\begin{equation}
I_0 = I_0\left(\frac{2 \alpha}{m a^2}\right), \quad
I_1 = I_1\left(\frac{2 \alpha}{m a^2}\right).
\end{equation}

Despite the fact that they do not move sideways, it is still interesting to 
investigate the spreading of minimal uncertainty wave packets on the lattice. 
In this case, the Green's function takes the form
\begin{equation}
G(x,t) = \frac{1}{2 \pi} \int_{- \pi/a}^{\pi/a} dp \ 
\exp\left(\frac{i \cos(p a) t}{m a^2} + i p x\right) =
\frac{1}{a} I_{x/a}\left(\frac{it}{m a^2}\right).
\end{equation}
Here $I_{x/a}$ is a modified Bessel function of degree $n = x/a \in \Z$. It 
should be noted that the Green's function is restricted to the lattice sites, 
i.e.\ $x = n a$ with $n \in \Z$. The spreading of a minimal uncertainty wave 
packet on the lattice, $\Phi(x,t) = A G(x,t - i \alpha)$, is illustrated in 
figure 2. Interestingly, the probability density shows an oscillatory behavior
which is absent in the continuum. This effect arises for wave packets of large 
energy that are sensitive to the lattice spacing scale $a$.
\begin{figure}[htb]
\begin{center}\hskip-8cm
\includegraphics[width=0.3\textwidth,angle=-90]{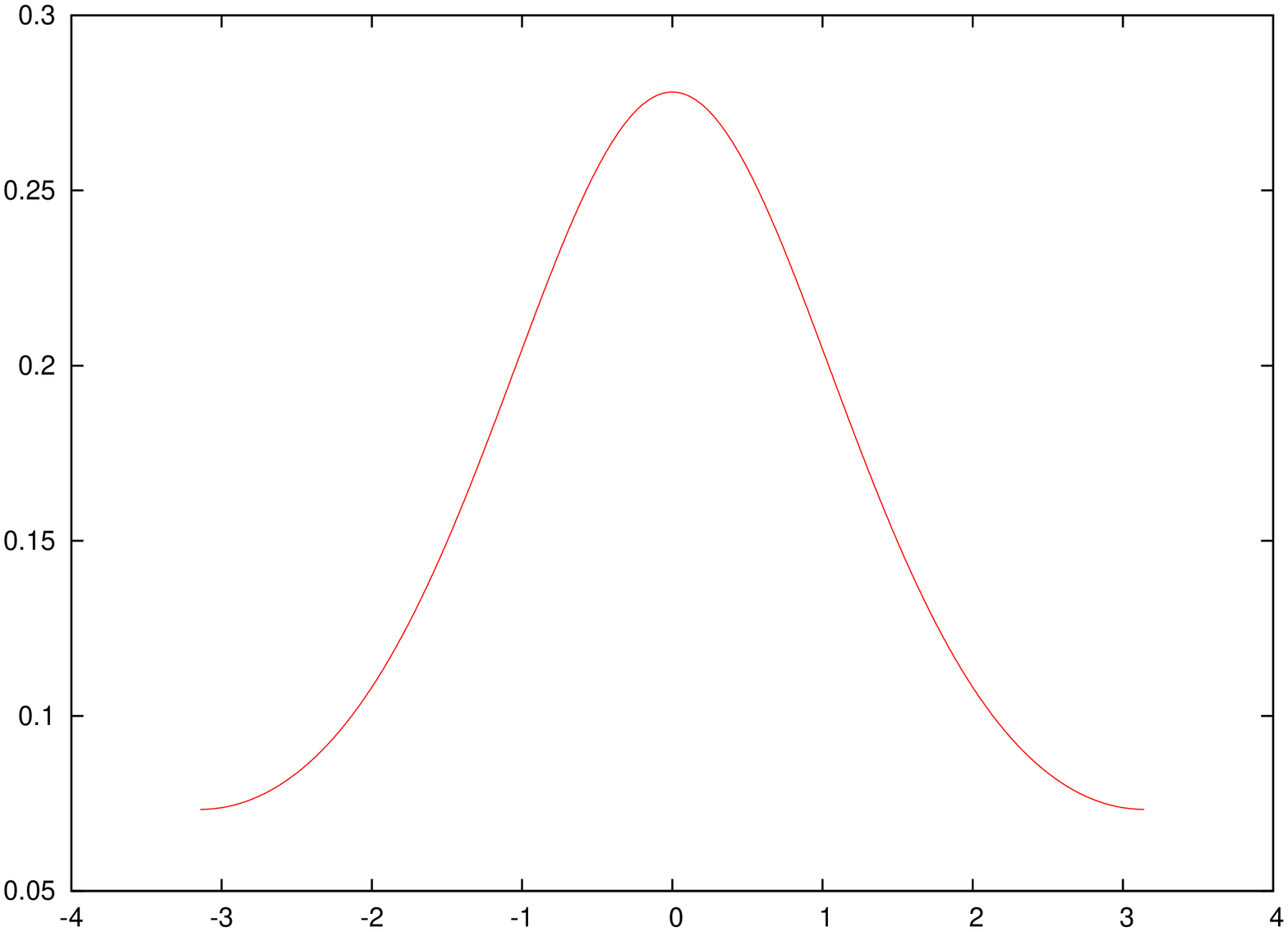}\vskip-5cm\hskip7cm
\includegraphics[width=0.52\textwidth]{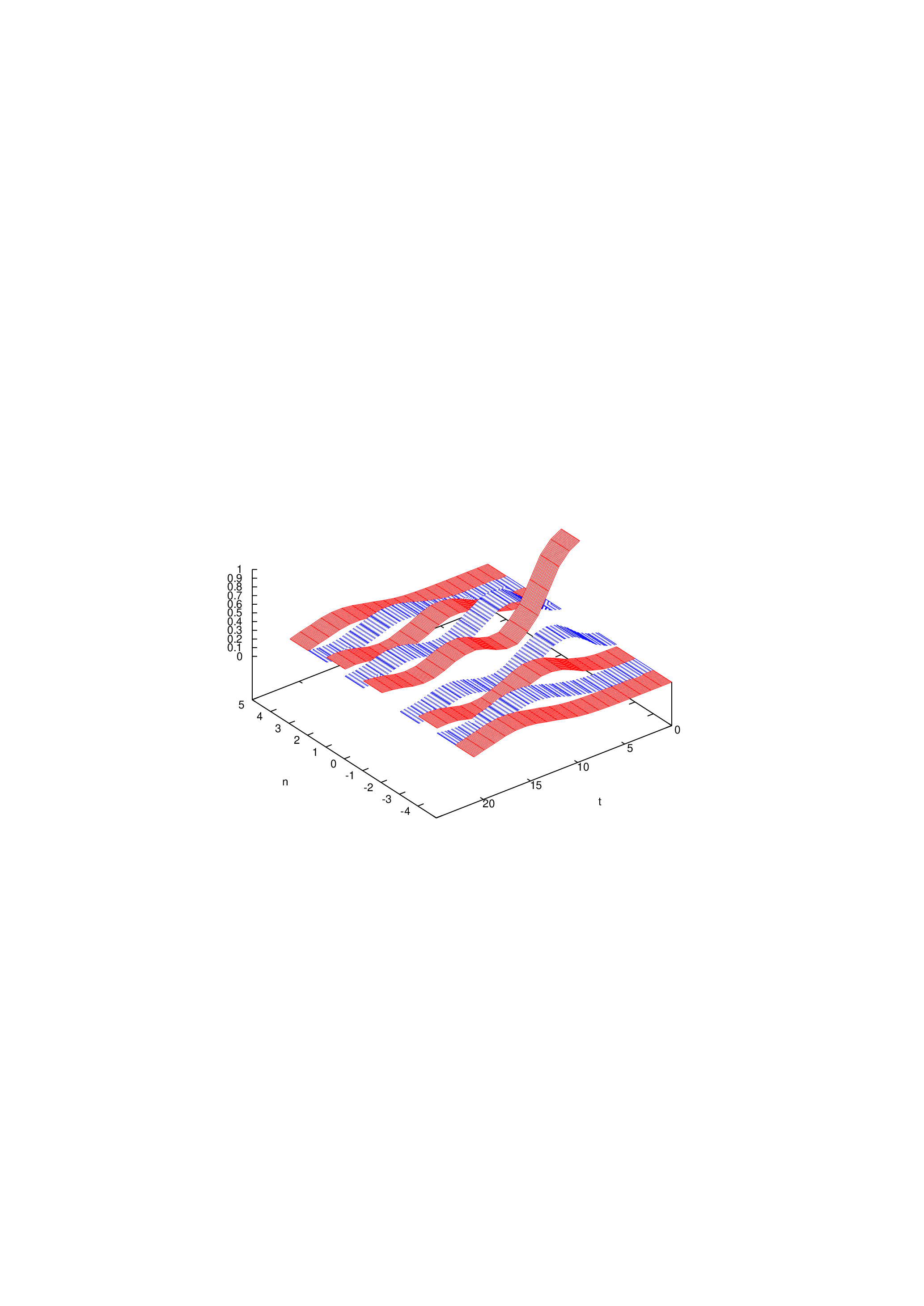}
\end{center}
\caption{\it Probability densities of a minimal position-velocity uncertainty 
wave packet on a lattice in momentum space (left) and spreading in coordinate
 space as a function of time (right) for $\alpha = 1$, $\beta = 0$ with $m=3$ 
and $a=1$.}
\end{figure}

\section{Spreading of Relativistic Wave Packets}

As we have stressed in the introduction, we do not consider the Dirac or 
Klein-Gordon equations because those belong to quantum field theory and not to 
relativistic quantum mechanics with a finite number of degrees of freedom
and a fixed number of particles. In this section we consider the spreading of
minimal uncertainty wave packets for a single free relativistic particle. The
spreading of other relativistic wave packets has been investigated in
\cite{Bak73,Alm84}. As a preparation, we first study the Lorentz transformation
properties of general wave functions.

\subsection{Poincar\'e Algebra and Boost Properties of General Wave Functions}

For a single free particle, it is trivial to satisfy the Poincar\'e algebra 
\begin{equation}
[H,P] = 0, \quad [M,P] = i H, \quad [M,H] = i P,
\end{equation}
by writing
\begin{equation}
H = \sqrt{p^2 + m^2}, \quad P = p, \quad 
M = \frac{1}{2}\left(x \sqrt{p^2 + m^2} + \sqrt{p^2 + m^2} \ x\right),
\end{equation}
for the Hamiltonian, the momentum, and the boost operator, respectively. The
unitary transformation that implements the boost to a frame moving with the
velocity $u$ is given by
\begin{equation}
U(u) = \exp(- i u M).
\end{equation}

Under the corresponding Lorentz transformation, the momentum $p$ of a particle
turns into
\begin{equation}
p' = \gamma \left[p - u E(p)\right] = 
\gamma \left[p - u \sqrt{p^2 + m^2} \right], \quad 
\gamma = \frac{1}{\sqrt{1 - u^2}}.
\end{equation}
Hence, the action of the boost operator on a plane wave is given by
\begin{equation}
U(u) \exp(i p x) = A(p) \exp(i p' x) = 
A(p) \exp\left(i \gamma \left[p - u E(p)\right]\right).
\end{equation}
The normalization condition
\begin{eqnarray}
\int dx \ A(p_1)^* \exp(- i p_1' x) A(p_2) \exp(i p_2' x)\!\!\!&=&\!\!\!
\int dx \ \exp(- i p_1 x) U(u)^\dagger U(u) \exp(i p_2 x) \nonumber \\
&=&\!\! 2 \pi \delta(p_1 - p_2),
\end{eqnarray}
then implies
\begin{equation}
\label{normA}
|A(p)|^2 = \gamma \left(1 - u \p_p E \right) = 
\gamma \left(1 - u v\right).
\end{equation}

When one applies the boost to an arbitrary wave packet
\begin{equation}
\Psi(x) = \frac{1}{2 \pi} \int dp \ \Psi(p) \exp\left(i p x\right),
\end{equation}
one hence obtains
\begin{eqnarray}
\label{Lorentzp}
\Psi_b(x)&=&U(u) \Psi(x) =
\frac{1}{2 \pi} \int dp \ \Psi(p) A(p) \exp\left(i p' x\right) \nonumber \\
&=&\frac{1}{2 \pi} \int dp' \ 
\gamma \left(1 + u v'\right) \Psi(p) A(p) \exp\left(i p' x\right).
\end{eqnarray}
Here we have used
\begin{equation}
p = \gamma\left[p' + u E'(p')\right], \quad 
\frac{dp}{dp'} = \gamma \left(1 + u \p_{p'} E' \right) =
\gamma \left(1 + u v'\right), \quad v' = \frac{v - u}{1 - u v},
\end{equation}
with $E'(p') = \sqrt{{p'}^2 + m^2}$. Using eq.(\ref{normA}) as well as 
eq.(\ref{Lorentzp}), one may then identify
\begin{equation}
\label{psiboost}
\Psi_b(p') = U(u) \Psi(p') =  A(-p') \Psi(p),
\end{equation}
with
\begin{equation}
|A(-p')|^2 = \gamma \left(1 + u \p_{p'} E' \right) = 
\gamma \left(1 + u v'\right).
\end{equation}
In momentum space, the norm of the boosted wave function then takes the form
\begin{eqnarray}
&&\frac{1}{2 \pi} \int dp' \ |\Psi_b(p')|^2 =
\frac{1}{2 \pi} \int dp' \ |A(-p')|^2 |\Psi(p)|^2 \nonumber \\
&&= \frac{1}{2 \pi} \int dp' \ \frac{dp}{dp'} |\Psi(p)|^2 = 
\frac{1}{2 \pi} \int dp \ |\Psi(p)|^2, 
\end{eqnarray}
which is thus indeed consistent.

Finally, in order to verify explicitly that $U(u) = \exp(- i u M)$, we expand 
for small $u$ and obtain
\begin{eqnarray}
U(u) - \1&=&- i u M + {\cal O}(u^2) = 
- i \frac{u}{2}\left(x \sqrt{p^2 + m^2} + \sqrt{p^2 + m^2} \ x\right) +
{\cal O}(u^2) \nonumber \\
&=&\frac{u}{2}\left(\p_p E(p) + E(p) \p_p\right) +{\cal O}(u^2) = 
\frac{u v}{2} + u E(p) \p_p + {\cal O}(u^2).
\end{eqnarray}
Acting with this operator on a momentum space wave function $\Psi(p')$ and
keeping only the leading linear order in $u$ we obtain
\begin{equation}
\left[U(u) - \1\right] \Psi(p') = 
\frac{u v}{2} \Psi(p) + u E(p) \p_p \Psi(p) + {\cal O}(u^2).
\end{equation}
Using eq.(\ref{psiboost}) and expanding for small $u$ in the same manner one
finds
\begin{eqnarray}
\left[U(u) - \1\right] \Psi(p')&=&A(-p') \Psi(p) - \Psi(p') \nonumber \\
&=&\left(1 + \frac{u v}{2}\right) \Psi(p) - \Psi(p) + u E(p) \p_p \Psi(p) + 
{\cal O}(u^2) \nonumber \\
&=&\frac{u v}{2} \Psi(p) + u E(p) \p_p \Psi(p) + {\cal O}(u^2),
\end{eqnarray}
which is thus indeed consistent. 

Using the boost properties of a general wave function, it is easy to show that
\begin{eqnarray}
&&\langle E \rangle_b = 
\gamma \left[\langle E \rangle - u \langle p \rangle\right], \quad
\langle p \rangle_b = 
\gamma \left[\langle p \rangle - u \langle E \rangle\right], \quad
\langle v \rangle_b = \langle v' \rangle =
\left\langle \frac{v - u}{1 - u v} \right\rangle, \nonumber \\
&&\langle x \rangle_b = \gamma \left\langle x + \frac{u}{2}
\left(v' x + x v'\right) \right\rangle, \quad 
\langle x^2 \rangle_b = \gamma^2 \left\langle 
\left[x + \frac{u}{2} \left(v' x + x v'\right)\right]^2 \right\rangle.
\end{eqnarray}
Here the subscript $b$ refers to expectation values taken with the boosted wave
function $\Psi_b$, while the expectation values without this subscript refer
to the original wave function $\Psi$.

\subsection{Relativistic Minimal Uncertainty Wave Packets}

According to the general expression of eq.(\ref{general}), the relativistic
minimal position-velocity uncertainty wave packets take the form
\begin{equation}
\Phi(p) = A \exp\left(- \alpha E(p) + \beta p\right) = 
A \exp\left(- \alpha \sqrt{p^2 + m^2} + \beta p\right).
\end{equation}
Similar wave packets have been discussed in the context of relativistic quantum
walks \cite{Str06}. Using eq.(\ref{psiboost}) it is easy 
to show that a boosted minimal uncertainty wave packet takes the form
\begin{equation}
\Phi_b(p') = U(u) \Phi(p') = A(-p') \Phi(p) = 
A(-p') A \exp(- \alpha' E(p') + \beta' p'),
\end{equation}
with
\begin{equation}
\alpha' = \gamma \left(\alpha - u \beta\right), \quad 
\beta' = \gamma \left(\beta - u \alpha\right),  
\end{equation}
i.e.\ $(\alpha,\beta)$ transforms as a space-time vector. However, due to the
factor $A(-p')$ (which is not constant), the boosted wave packet no longer has
minimal position-velocity uncertainty. This is in contrast to the 
non-relativistic case, in which a Galilean boost does not increase the
uncertainty product. In light of the relativistic position-velocity uncertainty
relation
\begin{equation}
\Delta x \Delta v \geq \frac{m^2}{2} \langle E^{-3} \rangle,
\end{equation}
this is not surprising because the uncertainty product does not transform
covariantly. We hence conclude that the concept of minimal position-velocity 
uncertainty is frame-dependent.

It is possible to work out the expectation values of a variety of operators for
relativistic wave packets with a minimal position-velocity uncertainty product.
As discussed in appendix A (for $\beta_i = 0$) one obtains
\begin{eqnarray}
&&\langle x \rangle = 0, \quad 
\langle x^2 \rangle = \alpha^2 - \beta^2 - 
\frac{2 \alpha \sqrt{\alpha^2 - \beta^2}}{K_1} 
\int_\alpha^\infty d\alpha' \ K_0\left(2 m \sqrt{{\alpha'}^2 - \beta^2}\right), 
\nonumber \\
&&\langle v \rangle = \frac{\beta}{\alpha}, \quad 
\langle v^2 \rangle = 1 - \frac{2 \sqrt{\alpha^2 - \beta^2}}{\alpha K_1} 
\int_\alpha^\infty d\alpha' \ K_0\left(2 m \sqrt{{\alpha'}^2 - \beta^2}\right),
\nonumber \\
&&\langle p \rangle = \frac{\beta}{\alpha^2 - \beta^2}
\left(1 + m \sqrt{\alpha^2 - \beta^2} \ \frac{K_0}{K_1} \right), \nonumber \\
&&\langle p^2 \rangle = \frac{m^2 \beta^2}{\alpha^2 - \beta^2} + 
\frac{\alpha^2 + 3 \beta^2}{2 (\alpha^2 - \beta^2)^2} 
\left(1 + m \sqrt{\alpha^2 - \beta^2} \ \frac{K_0}{K_1} \right),
\nonumber \\
&&\langle E \rangle = \frac{\alpha}{\alpha^2 - \beta^2} 
\left(1 + m \sqrt{\alpha^2 - \beta^2} \ \frac{K_0}{K_1} \right) - 
\frac{1}{2 \alpha}, \nonumber \\
&&\langle E^2 \rangle = \frac{m^2 \alpha^2}{\alpha^2 - \beta^2} +
\frac{\alpha^2 + 3 \beta^2}{2 (\alpha^2 - \beta^2)^2} 
\left(1 + m \sqrt{\alpha^2 - \beta^2} \ \frac{K_0}{K_1} \right).
\end{eqnarray}
Here $K_0$ and $K_1$ are modified Bessel functions of degree zero and one
\begin{equation}
K_0 = K_0\left(2 m \sqrt{\alpha^2 - \beta^2}\right), \quad
K_1 = K_1\left(2 m \sqrt{\alpha^2 - \beta^2}\right).
\end{equation}

\subsection{Relativistic Wave Packet Spreading and Apparent \\ Causality Violation}

Let us consider the Green's function in the relativistic case
\begin{equation}
G(x,t) = \frac{1}{2 \pi} \int dp \ \exp(- i \sqrt{p^2 + m^2} \ t + i p x). 
\end{equation}
Following methods presented in \cite{Sch89}, we now write
\begin{equation}
\sqrt{p^2 + m^2} = m \cosh z, \quad p = m \sinh z, \quad 
\frac{dp}{dz} = m \cosh z,
\end{equation}
as well as
\begin{eqnarray}
&&t = \sqrt{t^2 - x^2} \cosh\tau, \quad x = \sqrt{t^2 - x^2} \sinh\tau, \quad 
\mbox{for} \quad |x| < t, \nonumber \\
&&t = \sqrt{x^2 - t^2} \sinh\tau, \quad x = \sqrt{x^2 - t^2} \cosh\tau, \quad 
\mbox{for} \quad |x| > t,
\end{eqnarray}
such that
\begin{eqnarray}
\sqrt{p^2 + m^2} t - p x&=&m \sqrt{t^2 - x^2} 
\left(\cosh z \cosh\tau - \sinh z \sinh\tau\right) \nonumber \\
&=&m \sqrt{t^2 - x^2} \cosh(z - \tau), \quad 
\mbox{for} \quad |x| < t, \nonumber \\
\sqrt{p^2 + m^2} t - p x&=&m \sqrt{x^2 - t^2} 
\left(\cosh z \sinh\tau - \sinh z \cosh\tau\right) \nonumber \\
&=&m \sqrt{x^2 - t^2} \sinh(z - \tau), \quad \mbox{for} \quad |x| > t.
\end{eqnarray}
Inserting this in the expression for the Green's function we obtain
\begin{eqnarray}
G(x,t)&=&\frac{i}{2 \pi} \p_t \int dp \ \frac{1}{\sqrt{p^2 + m^2}}
\exp\left(- i \sqrt{p^2 + m^2} \ t + i p x\right) \nonumber \\
&=&\frac{i}{2 \pi} \p_t \int dz \ 
\exp\left(- i m \sqrt{t^2 - x^2} \cosh(z - \tau)\right) \nonumber \\
&=&\frac{i}{\pi} \p_t \int_0^\infty dz \ 
\left[\cos\left(m \sqrt{t^2 - x^2} \cosh z \right)\right. \nonumber \\
&-&\left.i \sin\left(m \sqrt{t^2 - x^2} \cosh z \right)\right], \quad 
\mbox{for} \quad |x| < t, \nonumber \\
G(x,t)&=&\frac{i}{2 \pi} \p_t \int dz \ 
\exp\left(- i m \sqrt{x^2 - t^2} \sinh(z - \tau)\right) \nonumber \\
&=&\frac{i}{\pi} \p_t \int_0^\infty dz \ 
\cos\left(m \sqrt{x^2 - t^2} \sinh z \right), \quad 
\mbox{for} \quad |x| > t.
\end{eqnarray}
Finally, using
\begin{eqnarray}
&&\int_0^\infty dz \ \sin\left(m \sqrt{t^2 - x^2} \cosh z \right) = 
\frac{\pi}{2} J_0\left(m \sqrt{t^2 - x^2}\right), \quad 
\mbox{for} \quad |x| < t, \nonumber \\
&&\int_0^\infty dz \ \cos\left(m \sqrt{t^2 - x^2} \cosh z \right) = 
- \frac{\pi}{2} N_0\left(m \sqrt{t^2 - x^2}\right), \quad 
\mbox{for} \quad |x| < t, \nonumber \\
&&\int_0^\infty dz \ \cos\left(m \sqrt{x^2 - t^2} \sinh z \right) = 
K_0\left(m \sqrt{x^2 - t^2}\right), \quad \mbox{for} \quad |x| > t.
\end{eqnarray}
where $J_0$, $N_0$, and $K_0$ are Bessel functions of degree zero, one obtains
\begin{eqnarray}
&&G(x,t) = \frac{1}{2} \p_t \left[J_0\left(m \sqrt{t^2 - x^2}\right) - 
i N_0\left(m \sqrt{t^2 - x^2}\right)\right] \quad \mbox{for} \quad |x| < t, 
\nonumber \\
&&G(x,t) = \frac{i}{\pi} \p_t K_0\left(m \sqrt{x^2 - t^2}\right) \quad 
\mbox{for} \quad |x| > t.
\end{eqnarray}
Using $\frac{i}{\pi} K_0(i z) = \frac{1}{2}[J_0(z) - i N_0(z)]$, one can write  
\begin{equation}
G(x,t) = \frac{i}{\pi} \p_t K_0\left(m \sqrt{x^2 - t^2}\right) =
- \frac{i m t}{\pi \sqrt{x^2 - t^2}} K_1\left(m \sqrt{x^2 - t^2}\right),
\end{equation}
for all values of $x$ and $t$. Here $K_1$ is a modified Bessel function of 
degree one.

Interestingly, the Green's function does not vanish at space-like distances
$|x| > t$, which seems to violate causality \cite{Fle65}. As discussed in 
appendix B, the apparent violation of causality is resolved in the framework of
quantum field theory, and is due to an inherent non-locality of single particle
states \cite{Ree61,Fle65,Heg74,Heg80,Heg85,Ros87,Mos90,Heg01,Bar03,Bus05,Bus06}.
The probability density of a minimal position-velocity uncertainty wave packet 
given by $\Phi(x,t) = A G(x - i \beta,t - i \alpha)$ is illustrated in figure 3.
\begin{figure}[htb]
\begin{center}
\includegraphics[width=0.3\textwidth,angle=-90]{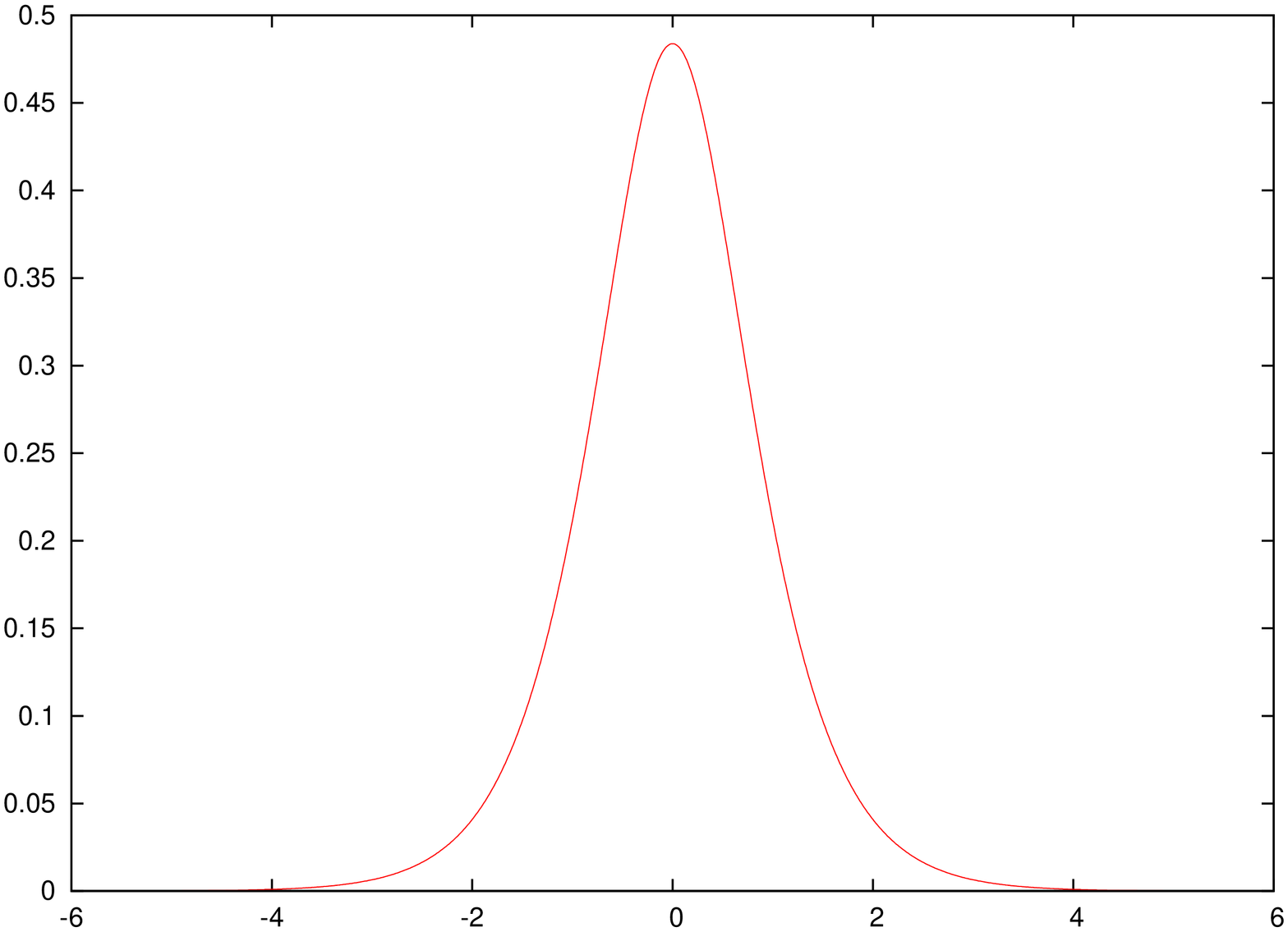} \hskip-0.5cm
\includegraphics[width=0.41\textwidth,angle=-90]{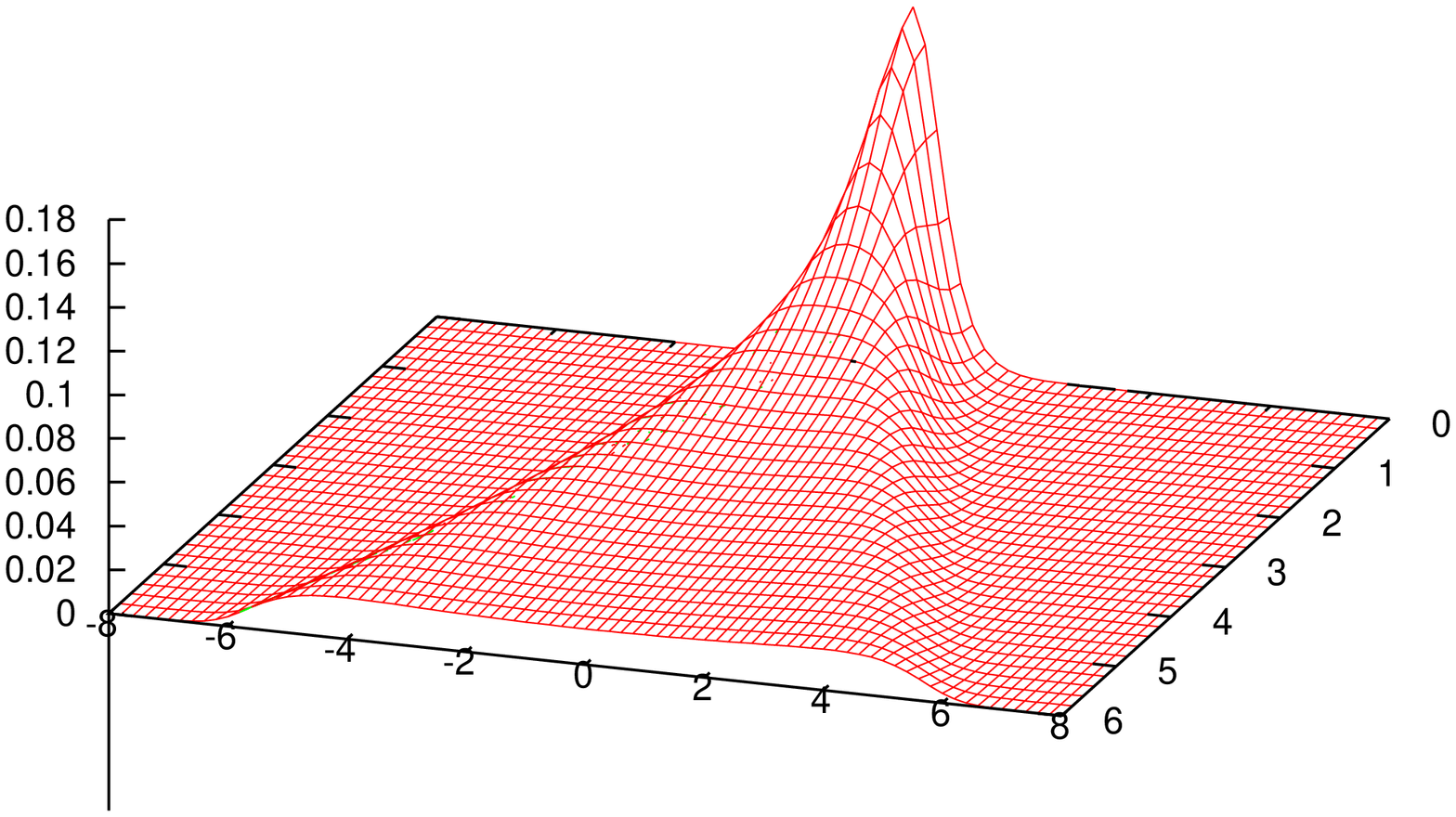}
\includegraphics[width=0.3\textwidth,angle=-90]{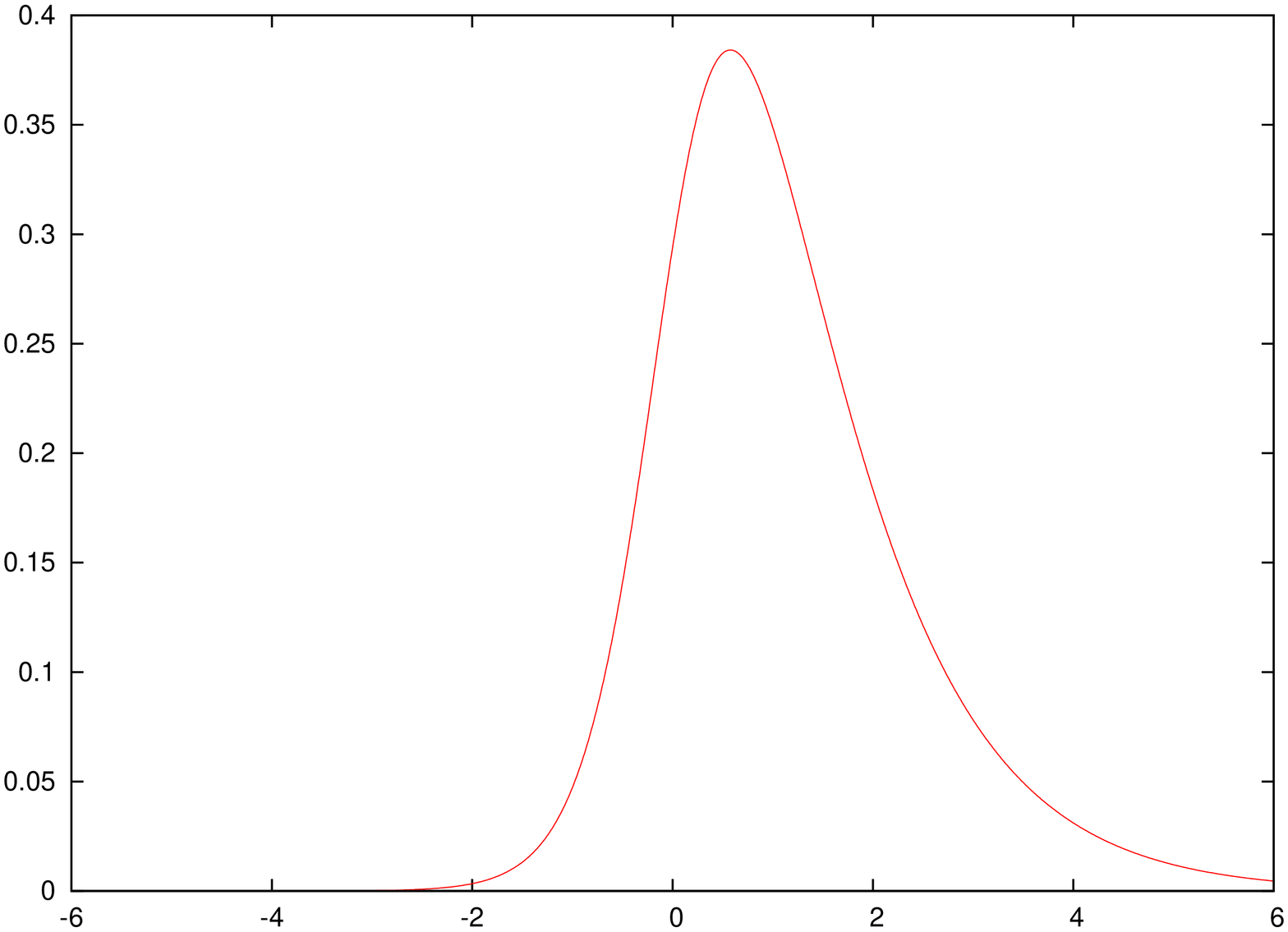} \hskip-0.5cm
\includegraphics[width=0.41\textwidth,angle=-90]{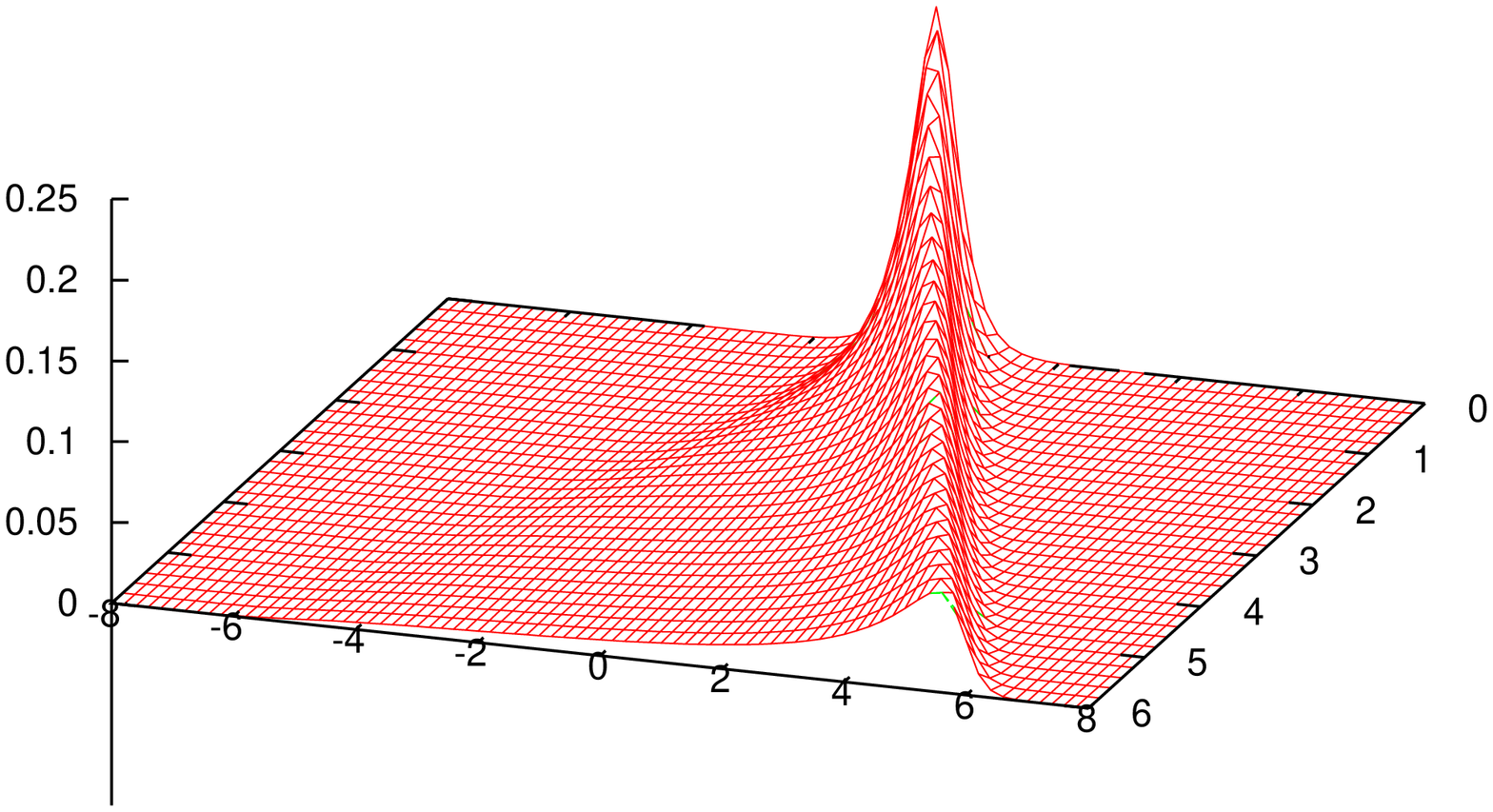}
\vskip-1cm
\end{center}\vskip-1cm
\caption{\it Probability densities of two relativistic minimal position-velocity
uncertainty wave packets in momentum space (left) and spreading in coordinate 
space as a function of time (right) for $\alpha = 1$, $\beta = 0$ (top) and for
$\alpha = 1$, $\beta = 1/2$ (bottom) with $m=1$.}
\end{figure}

\subsection{The Massless Limit}

It is interesting to consider the massless limit $m \rightarrow 0$. A minimal
position-velocity uncertainty wave packet then takes the form
\begin{equation}
\Phi(p) = A \exp(- \alpha |p| + \beta p). 
\end{equation}
It is straightforward to work out the expectation values of various operators 
and (for $\beta_i = 0$) one obtains
\begin{eqnarray}
&&\langle x \rangle = 0, \quad \langle x^2 \rangle = (\Delta x)^2 = 
\alpha^2 - \beta^2, \nonumber \\
&&\langle v \rangle = \frac{\beta}{\alpha}, \quad \langle v^2 \rangle = 1, 
\quad (\Delta v)^2 = 1 - \frac{\beta^2}{\alpha^2}, \nonumber \\
&&\langle p \rangle = \frac{\beta}{\alpha^2 - \beta^2}, \quad 
\langle p^2 \rangle = \frac{\alpha^2 + 3 \beta^2}{2 (\alpha^2 - \beta^2)^2}, 
\quad (\Delta p)^2 = \frac{\alpha^2 + \beta^2}{2 (\alpha^2 - \beta^2)^2},
\nonumber \\
&&\langle E \rangle = 
\frac{\alpha^2 + \beta^2}{2 \alpha (\alpha^2 - \beta^2)}, \quad 
\langle E^2 \rangle = \frac{\alpha^2 + 3 \beta^2}{2 (\alpha^2 - \beta^2)^2}, 
\quad (\Delta E)^2 = \frac{\alpha^4 + 4 \alpha^2 \beta^2 - \beta^4}
{4 \alpha^2 (\alpha^2 - \beta^2)^2}. \nonumber \\ \
\end{eqnarray}
It should be noted that, despite the fact that massless particles move with the
speed of light, in general $|\langle v \rangle| \leq 1$, because a particle may
move simultaneously both to the left and to the right with non-zero probability
amplitude. Only for $|\beta| \rightarrow \alpha$ the particle is entirely left-
or right-moving, $|\langle v \rangle| \rightarrow 1$, $\Delta v \rightarrow 0$, 
and the corresponding wave packet is not spreading.

In the massless limit, the Green's function takes the form
\begin{equation}
G(x,t) = \frac{1}{2 \pi} \int dp \ \exp(- i |p| t + i p x) =  
\frac{i}{\pi} \frac{t}{x^2 - t^2}.
\end{equation}
Accordingly, the wave function of a minimal uncertainty wave packet is given by 
$\Phi(x,t) = A G(x - i \beta,t - i \alpha)$. The time-dependence of two
corresponding probability densities is shown in figure 4.
\begin{figure}[htb]
\begin{center}
\includegraphics[width=0.3\textwidth,angle=-90]{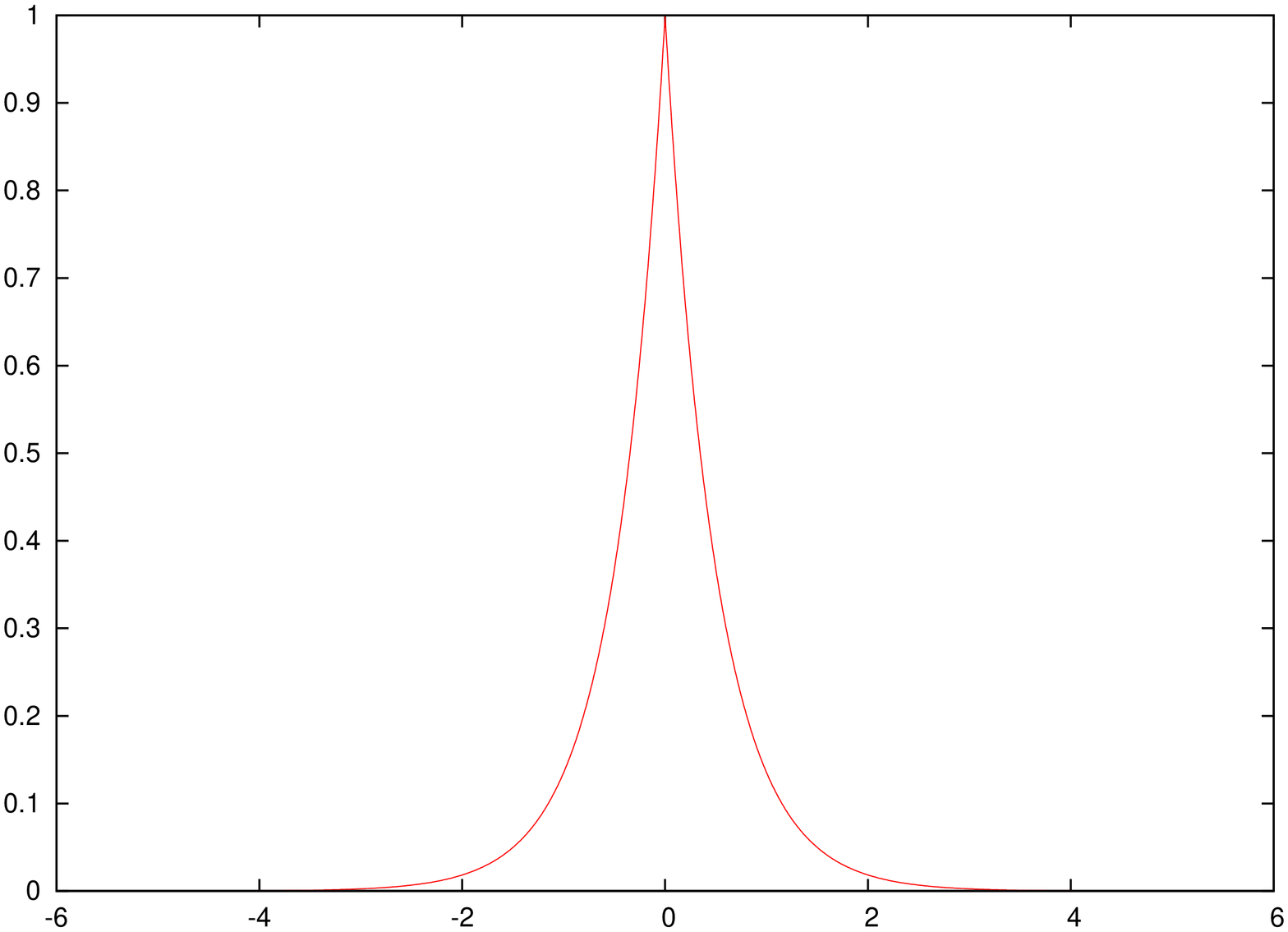} \hskip-0.5cm
\includegraphics[width=0.41\textwidth,angle=-90]{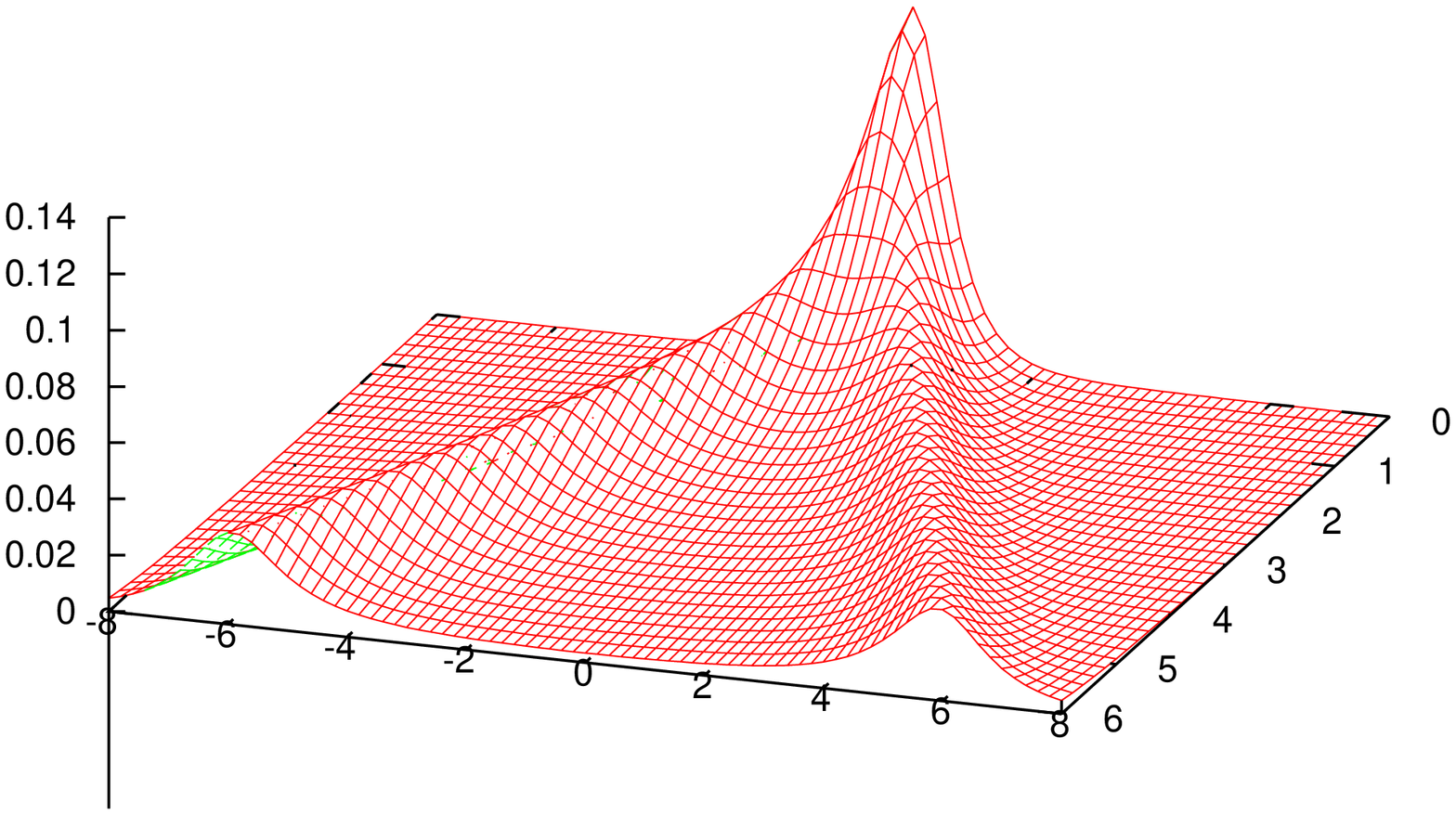}
\includegraphics[width=0.3\textwidth,angle=-90]{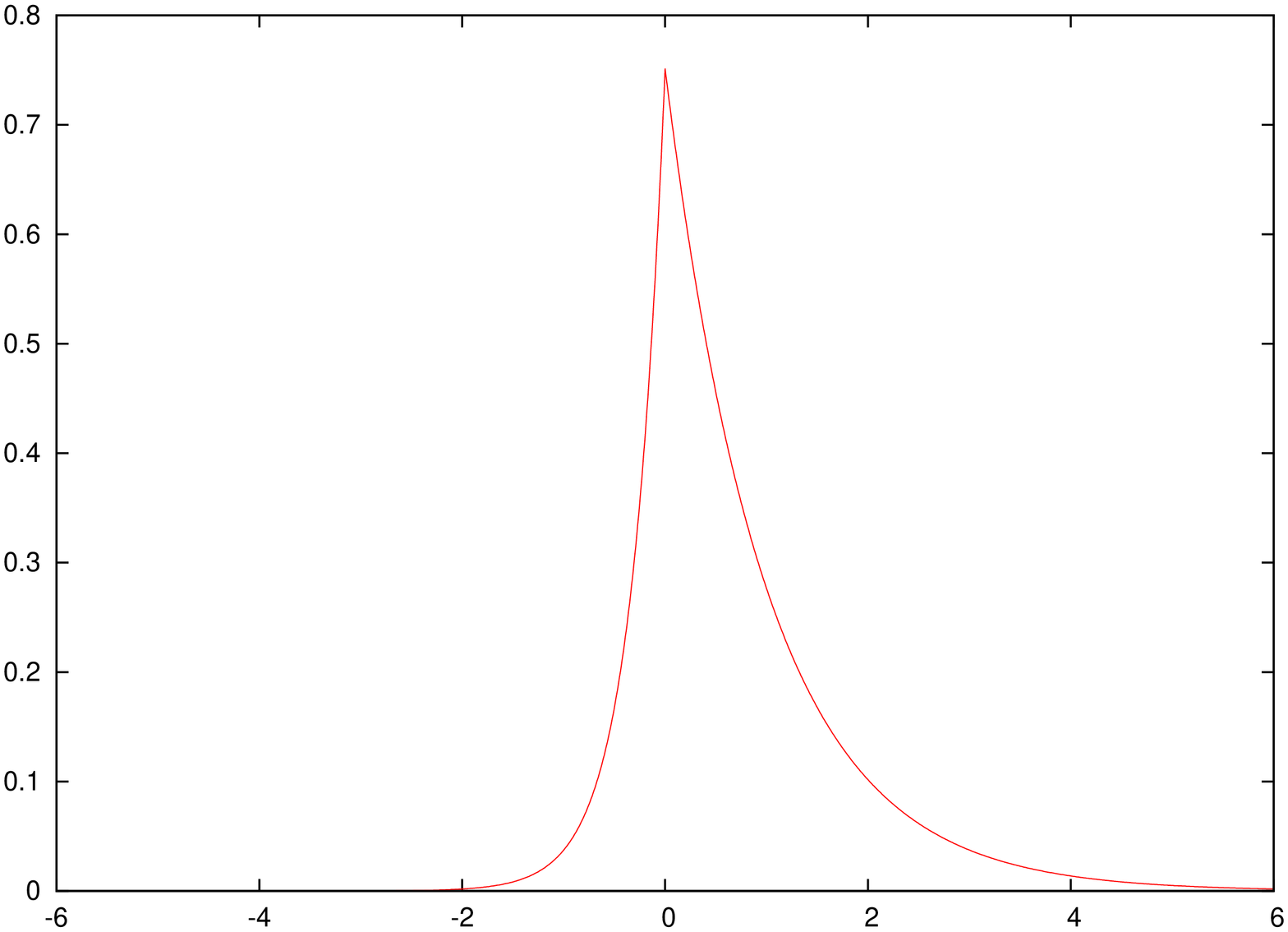} \hskip-0.5cm
\includegraphics[width=0.41\textwidth,angle=-90]{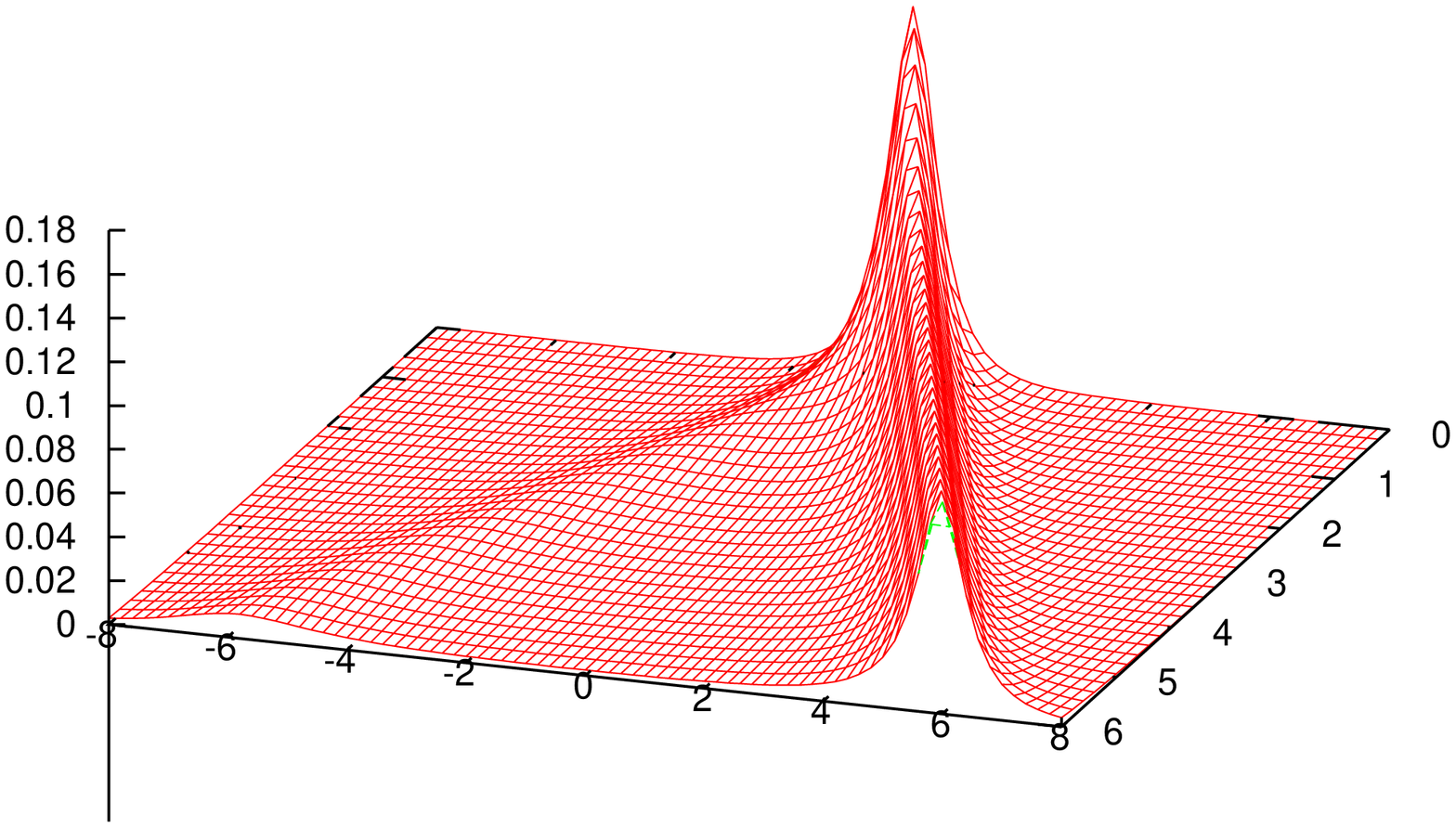}
\vskip-1cm
\end{center}\vskip-1cm
\caption{\it Probability densities of two ultra-relativistic minimal 
position-velocity uncertainty wave packets in momentum space (left) and 
spreading in coordinate space as a function of time (right) for $\alpha = 1$, 
$\beta = 0$ (top) and for $\alpha = 1$, $\beta = 1/2$ (bottom) with $m=0$.}
\end{figure}
In contrast to the non-relativistic case, the spreading of ultra-relativistic 
wave packets proceeds by wave packet splitting into two packets, one moving to 
the left and one moving to the right, each with the speed of light.

\section{Propagation of Wave Packets in an Expanding Universe}

In this section we consider the propagation of wave packets in an expanding 
Universe. For simplicity, we limit ourselves to one spatial dimension, but the 
generalization to higher dimensions is straightforward.

\subsection{Free Falling Particle in an Expanding Universe}

Let us consider an expanding 1-dimensional Universe with the 
Freedman-Lemaitre-Robertson-Walker-type metric
\begin{equation}
(ds)^2 = (dt)^2 - R(t)^2 (d\rho)^2.
\end{equation}
Here $R(t)$ is the scale parameter of the Universe whose time-dependence we 
consider as given. The position $x = R(t) \rho$ of a particle is described by 
the dimensionless coordinate $\rho$. The Lagrange function of a free falling
particle then takes the form
\begin{equation}
L = - m \frac{ds}{dt} =  - m \sqrt{1 - R(t)^2 \dot\rho^2},
\end{equation}
and the momentum canonically conjugate to the dimensionless coordinate $\rho$ is
given by
\begin{equation}
p_\rho = \frac{\p L}{\p \dot\rho} = \frac{m R(t)^2 \dot\rho}
{\sqrt{1 - R(t)^2 \dot\rho^2}},
\end{equation}
while the dimensionful momentum conjugate to $x$ is $p = p_\rho/R(t)$. The 
corresponding time-dependent classical Hamilton function hence takes the form
\begin{equation}
H(t) = p_\rho \dot\rho - L = \sqrt{p_\rho^2/R(t)^2 + m^2}.
\end{equation}
The classical equations of motion are thus given by
\begin{equation}
\dot p_\rho = - \frac{\p H}{\p \rho} = 0, \quad 
\dot \rho = \frac{\p H}{\p p_\rho} = 
\frac{p_\rho/R(t)^2}{\sqrt{p_\rho^2/R(t)^2 + m^2}}.
\end{equation}
Integrating the two equations one obtains
\begin{equation}
\rho(t) = \rho(0) + \int_0^t dt' \ 
\frac{p_\rho/R(t')^2}{\sqrt{p_\rho^2/R(t')^2 + m^2}}.
\end{equation}
The corresponding equation for the particle's position then takes the form
\begin{equation}
x(t) = R(t) \rho(t) = \frac{R(t)}{R(0)} x(0) + 
R(t) \int_0^t dt' \ \frac{v(t')}{R(t')},
\end{equation}
where we have identified the velocity as
\begin{equation}
v(t) = \frac{p_\rho/R(t)}{\sqrt{p_\rho^2/R(t)^2 + m^2}}.
\end{equation}
In particular, as the Universe expands, the velocity $v(t)$ decreases, because
the momentum $p = p_\rho/R(t)$ is red-shifted. The time-dependence of the
velocity is given by
\begin{equation}
\label{velocity}
v(t) = \frac{R(0)}{R(t)} \frac{v(0)}{\sqrt{1 - v(0)^2 + v(0)^2 R(0)^2/R(t)^2}}.
\end{equation}

Upon canonical quantization the Hamilton function turns into a Hamilton 
operator, and one postulates $[\rho,p_\rho] = i$, which is realized by 
$\rho = i \p_{p_\rho}$. The Schr\"odinger equation then takes the form
\begin{equation}
i \p_t \Psi(p_\rho,t) = H(t) \Psi(p_\rho,t),
\end{equation}
which is solved by
\begin{equation}
\Psi(p_\rho,t) = 
\exp\left(- i \int dt \ \sqrt{p_\rho^2/R(t)^2 + m^2} \right) 
\Psi(p_\rho),
\end{equation}
where $\Psi(p_\rho)$ is the momentum space wave function at $t = 0$.

\subsection{Time-dependence of Expectation Values}

Let us consider the time-dependence of the expectation value of the 
dimensionless coordinate
\begin{eqnarray}
\langle \rho \rangle(t)&=&\frac{1}{2 \pi} \int dp_\rho \
\Psi(p_\rho,t)^* i \p_{p_\rho} \Psi(p_\rho,t) \nonumber \\
&=&\langle \rho \rangle(0) + 
\int_0^t dt' \ \frac{1}{2 \pi} \int dp_\rho \ |\Psi(p_\rho)|^2
\frac{p_\rho/R(t')^2}{\sqrt{p_\rho^2/R(t')^2 + m^2}} \nonumber \\
&=&\langle \rho \rangle(0) + 
\int_0^t dt' \ \frac{\langle v(t') \rangle}{R(t')}.
\end{eqnarray}
Here the expectation value of the velocity $v(t')$ is evaluated for the initial
wave function $\Psi(p_\rho)$. Correspondingly, one obtains
\begin{equation}
\langle x \rangle(t) = \frac{R(t)}{R(0)} \langle x \rangle(0) + 
R(t) \int_0^t dt' \ \frac{\langle v(t') \rangle}{R(t')}.
\end{equation}
Similarly, one finds
\begin{equation}
\langle \rho^2 \rangle(t) = \langle \rho^2 \rangle(0) +
\int_0^t dt' \ \frac{1}{R(t')} \langle v(t') \rho + \rho v(t') \rangle +
\left\langle \left(\int_0^t dt' \ \frac{1}{R(t')} v(t') \right)^2 \right\rangle.
\end{equation}
From this it is straightforward to obtain an expression for $\Delta x(t)$.

\subsection{Propagation of Minimal Uncertainty Wave Packets in an Expanding
Universe}

For a relativistic minimal position-velocity uncertainty wave packet the 
initial wave function at $t = 0$ is given by
\begin{equation}
\Psi(p_\rho) = 
A \exp\left(- \alpha \sqrt{p_\rho^2/R(0)^2 + m^2} + \beta p_\rho/R(0)\right),
\end{equation}
and the velocity expectation value takes the form
\begin{eqnarray}
\langle v(t) \rangle&=&\frac{1}{2 \pi} \int dp_\rho \ |\Psi(p_\rho)|^2 
\frac{p_\rho/R(t)}{\sqrt{p_\rho^2/R(t)^2 + m^2}} \nonumber \\
&=&\frac{R(0)}{R(t)} \frac{|A|^2 R(0)}{2 \pi} \int dp \ 
\exp\left(- 2 \alpha \sqrt{p^2 + m^2} + 2 \beta p \right)
\frac{p}{\sqrt{p^2 \frac{R(0)^2}{R(t)^2} + m^2}}. \nonumber \\ \
\end{eqnarray}
We have not been able to simplify this integral any further. However, in the
massless case it simplifies to
\begin{equation}
\langle v(t) \rangle = \frac{1}{2 \pi} \int dp_\rho \ |\Psi(p_\rho)|^2 
\mbox{sign}(p_\rho) = \langle v(0) \rangle = \frac{\beta}{\alpha},
\end{equation}
i.e.\ the average velocity is not red-shifted. One should keep in mind that
$\langle v(0) \rangle$ receives contributions $\pm 1$, corresponding to the
massless particle traveling to the left or to the right with the velocity of 
light. Similarly, in the non-relativistic limit 
\begin{eqnarray}
\langle v(t) \rangle&=&\frac{1}{2 \pi} \int dp_\rho \ |\Psi(p_\rho)|^2 
\frac{p_\rho}{m R(t)} \nonumber \\
&=&\frac{R(0)}{R(t)} \frac{|A|^2 R(0)}{2 \pi} \int dp \ 
\exp\left(- 2 \alpha \left(m + p^2/2m\right) + 2 \beta p \right) \frac{p}{m}. 
\nonumber \\
&=&\frac{R(0)}{R(t)} \langle v(0) \rangle = 
\frac{R(0)}{R(t)} \frac{\beta}{\alpha},
\end{eqnarray}
i.e.\ the velocity is red-shifted in proportion to the scale parameter.

\section{Conclusions}

While most results obtained in this paper are rather simple, except for a few,
we have not been able to find them in the physics literature. The standard 
textbook example of a spreading Gaussian wave packet is just the simplest case 
of a minimal position-velocity wave packet, which can be defined for an 
arbitrary relativistic or non-relativistic dispersion relation $E(p)$. Such wave
packets saturate a generalized position-velocity uncertainty relation, and their
time-evolution is described by analytic continuation of the corresponding 
Green's function. Detailed analytic solutions have been worked out for a
non-relativistic particle in the continuum and on the lattice as well as for a
relativistic particle, both in the massive and in the massless case.

Some of our results belong to
relativistic quantum mechanics. Of course, since quantum field theory has 
been identified as the correct description of Nature at the most fundamental 
level that is accessible today, there is no urgent need for relativistic quantum
mechanics. In particular, in view of Leutwyler's no-interaction theorem
\cite{Leu65,Cur63}, relativistic quantum mechanics seems to be limited to free 
theories, although we have already mentioned \cite{Rui80,Rui86,Rui01} to which 
the theorem does not apply. In fact, there may exist further interacting systems
with a fixed number of particles in relativistic quantum mechanics, for 
example, with contact interactions. Even if no further systems of this kind 
should exist, we believe that the results presented here may be of some value. 
In particular, they may help bridging the gap between non-relativistic 
quantum mechanics and relativistic quantum field theory, which makes learning 
the latter rather non-trivial. The explicit solutions of spreading relativistic
wave packets illustrate in a simple setting what happens when both relativistic
and quantum effects are present at the same time.

Although we have not elaborated on this, we can imagine that our results may 
have some use in neutrino physics. Indeed, the spreading of neutrino wave
packets has been discussed in various places in the literature 
\cite{Giu91,Giu04}, mostly using Gaussian wave packets. As we have discussed, 
Gaussian wave packets are natural to consider in non-relativistic quantum 
mechanics. In relativistic theories, on the other hand, the minimal 
position-velocity wave packets discussed above seem more natural, in 
particular, since closed analytic expressions have been obtained for a large 
variety of observables. Hence, in the relativistic case, there is no need to 
use Gaussian wave packets, which only yield approximate analytic results. 
Whether wave packet spreading of neutrinos (or other light particles) either in
a static or in an expanding Universe is related to phenomenologically relevant 
questions remains an interesting topic for future investigations.

\section*{Acknowledgements}

We like to thank J.\ Balog and H.\ Leutwyler for interesting discussions.
This work is supported in parts by the Schweizerischer Nationalfonds (SNF) as 
well as by the Swiss national qualification program 
Bio\-medi\-zin-Na\-tur\-wis\-sen\-schaft-For\-schung (BNF).

\begin{appendix}

\section{Evaluation of Various Expectation Values}

In this appendix we work out the expectation values of various operators, for
minimal position-velocity uncertainty wave packets both on the lattice and in
the relativistic case.

\subsection{Expectation Values for Wave Packets on the Lattice}

In this subsection we calculate expectation values for the wave packet
\begin{equation}
\Phi(p) = A \exp\left(\frac{\alpha \cos(p a)}{m a^2} \right).
\end{equation}
The normalization condition then takes the form
\begin{equation}
\frac{1}{2 \pi} \int dp \ |\Phi(p)|^2 = \frac{|A|^2}{2 \pi} \int dp \
\exp\left(\frac{2 \alpha \cos(p a)}{m a^2}\right) = \frac{|A|^2}{a}
I_0 = 1 \ \Rightarrow \ |A|^{-2} = \frac{1}{a} I_0.
\end{equation}
For the energy one obtains
\begin{eqnarray}
\langle E \rangle&=&
- \frac{1}{2 \pi} \int dp \ |\Phi(p)|^2 \ \frac{\cos(pa)}{m a^2} \nonumber \\
&=&- \frac{|A|^2}{2 \pi} \int dp \
\exp\left(\frac{2 \alpha \cos(p a)}{m a^2}\right) \frac{\cos(pa)}{m a^2} 
\nonumber \\
&=&- \frac{|A|^2}{m a^3} I_1 = - \frac{I_1}{I_0 m a^2},
\end{eqnarray}
where $I_0$ and $I_1$ are modified Bessel functions of degree zero and one
\begin{equation}
I_0 = I_0\left(\frac{2 \alpha}{m a^2}\right), \quad
I_1 = I_1\left(\frac{2 \alpha}{m a^2}\right).
\end{equation}
For the energy squared we find
\begin{eqnarray}
\langle E^2 \rangle&=&
\frac{1}{2 \pi} \int dp \ |\Phi(p)|^2 \ \frac{\cos^2(pa)}{m^2 a^4} 
\nonumber \\
&=&\frac{|A|^2}{2 \pi} \int dp \
\exp\left(\frac{2 \alpha \cos(p a)}{m a^2}\right) \frac{\cos^2(pa)}{m^2 a^4} 
\nonumber \\
&=&\frac{|A|^2}{4 a} \p_\alpha^2 I_0 = \frac{\p_\alpha^2 I_0}{4 I_0}
= \frac{1}{m^2 a^4} \left(1 - \frac{m a^2 I_1}{2 \alpha I_0}\right).
\end{eqnarray}
The expectation value of the velocity squared takes the form
\begin{equation}
\langle v^2 \rangle = \left\langle \frac{\sin^2(pa)}{m^2 a^2} \right\rangle =
\frac{1}{m^2 a^2} - a^2 \langle E^2 \rangle = \frac{I_1}{2 m \alpha I_0}.
\end{equation}
Finally, we consider
\begin{eqnarray}
\langle x^2 \rangle&=&- \frac{1}{2 \pi} \int dp \ \Phi(p)^* \p_p^2 \Phi(p) =
\left \langle \frac{\alpha \cos(pa)}{m} -
\frac{\alpha^2 \sin^2(pa)}{m^2 a^2}\right\rangle \nonumber \\
&=&- \alpha a^2 \langle E \rangle - \alpha^2 \langle v^2 \rangle =
\frac{\alpha I_1}{2 m I_0}.
\end{eqnarray}

\subsection{Expectation Values for Relativistic Wave Packets}

In this subsection we calculate expectation values for the minimal
position-velocity uncertainty wave packet
\begin{equation}
\Phi(p) = A \exp\left(- \alpha \sqrt{p^2 + m^2} + \beta p\right).
\end{equation}
Here we limit ourselves to $\beta \in \R$ which correspond to a wave packet 
centered at $x = 0$. Adding an imaginary part to $\beta$ leads to a simple
translation of the wave packet. The normalization condition takes the form
\begin{equation}
\frac{1}{2 \pi} \int dp \ |\Phi(p)|^2 = \frac{|A|^2}{2 \pi} \int dp \
\exp\left(- 2 \alpha \sqrt{p^2 + m^2} + 2 \beta p\right) = 1.
\end{equation}
Introducing
\begin{eqnarray}
&&\sqrt{p^2 + m^2} = m \cosh z, \quad p = m \sinh z, \quad \frac{dp}{dz} =
m \cosh z, \nonumber \\
&&\alpha = \sqrt{\alpha^2 - \beta^2} \cosh\lambda, \quad 
\beta = \sqrt{\alpha^2 - \beta^2} \sinh\lambda, 
\end{eqnarray}
one obtains
\begin{equation}
\alpha \sqrt{p^2 + m^2} - \beta p = \cosh z \cosh\lambda - \sinh z \sinh\lambda
= \cosh(z - \lambda),
\end{equation}
and the normalization condition thus takes the form
\begin{equation}
\frac{|A|^2 m}{2 \pi} \int dz \ \cosh z
\exp\left(- 2 m \sqrt{\alpha^2 - \beta^2} \cosh(z - \lambda)\right) = 1.
\end{equation}
Let us first consider the integral
\begin{eqnarray}
&&\frac{1}{2 \pi} 
\int dz \ \exp\left(- 2 \alpha m \cosh z + 2 \beta m \sinh z\right) = 
\nonumber \\ 
&&\frac{1}{2 \pi} \int dz \ \exp\left(- 2 m \sqrt{\alpha^2 - \beta^2} \cosh z 
\right) = \frac{1}{\pi} K_0.
\end{eqnarray}
The normalization condition can now be expressed as
\begin{equation}
- \frac{|A|^2}{2 \pi} \p_\alpha K_0
= \frac{|A|^2 m \alpha}{\pi \sqrt{\alpha^2 - \beta^2}} K_1 = 1 \ \Rightarrow \ 
|A|^{-2} = \frac{m \alpha}{\pi \sqrt{\alpha^2 - \beta^2}} K_1. 
\end{equation}
Here $K_0$ and $K_1$ are modified Bessel functions of degree zero and one
\begin{equation}
K_0 = K_0\left(2 m \sqrt{\alpha^2 - \beta^2}\right), \quad
K_1 = K_1\left(2 m \sqrt{\alpha^2 - \beta^2}\right).
\end{equation}

Next we consider the expectation value of the velocity
\begin{eqnarray}
\langle v \rangle&=&
\frac{1}{2 \pi} \int dp \ |\Phi(p)|^2 \frac{p}{\sqrt{p^2 + m^2}} \nonumber \\
&=&\frac{|A|^2 m}{2 \pi} \int dz \ \sinh z
\exp\left(- 2 \alpha m \cosh z + 2 \beta m p \sinh z\right) \nonumber \\
&=&\frac{|A|^2}{2 \pi} \p_\beta K_0 = - \frac{\p_\beta K_0}{\p_\alpha K_0} = 
\frac{\beta}{\alpha}.
\end{eqnarray}
Similarly, we obtain
\begin{eqnarray}
\langle v^2 \rangle&=&
\frac{1}{2 \pi} \int dp \ |\Phi(p)|^2 \frac{p^2}{p^2 + m^2} \nonumber \\
&=&1 - \frac{|A|^2 m}{2 \pi} \int dz \ \frac{1}{\cosh z}
\exp\left(- 2 \alpha m \cosh z + 2 \beta m p \sinh z\right) \nonumber \\
&=&1 - \frac{2 |A|^2 m}{\pi} \int_\alpha^\infty d\alpha' \ 
K_0\left(2 m \sqrt{{\alpha'}^2 - \beta^2}\right) \nonumber \\
&=&1 - \frac{2 \sqrt{\alpha^2 - \beta^2}}{\alpha K_1} 
\int_\alpha^\infty d\alpha' \ K_0\left(2 m \sqrt{{\alpha'}^2 - \beta^2}\right),
\end{eqnarray}
which he have not been able to simplify further. Next, we consider 
\begin{eqnarray}
\langle p \rangle&=&
\frac{1}{2 \pi} \int dp \ |\Phi(p)|^2 p \nonumber \\
&=&\frac{|A|^2 m^2}{2 \pi} \int dz \ \cosh z \sinh z
\exp\left(- 2 \alpha m \cosh z + 2 \beta m p \sinh z\right) \nonumber \\
&=&- \frac{|A|^2}{4 \pi} \p_\alpha \p_\beta K_0 =
\frac{\p_\alpha \p_\beta K_0}{2 \p_\alpha K_0} =
\frac{\beta}{\alpha^2 - \beta^2}
\left(1 + m \sqrt{\alpha^2 - \beta^2} \ \frac{K_0}{K_1} \right).
\end{eqnarray}
Similarly, we obtain
\begin{eqnarray}
\langle p^2 \rangle&=&
\frac{1}{2 \pi} \int dp \ |\Phi(p)|^2 p^2 \nonumber \\
&=&\frac{|A|^2 m^3}{2 \pi} \int dz \ \cosh z \sinh^2 z
\exp\left(- 2 \alpha m \cosh z + 2 \beta m p \sinh z\right) \nonumber \\
&=&- \frac{|A|^2}{8 \pi} \p_\alpha \p_\beta^2 K_0 = 
\frac{\p_\alpha \p_\beta^2 K_0}{4 \p_\alpha K_0} \nonumber \\
&=&\frac{m^2 \beta^2}{\alpha^2 - \beta^2} + 
\frac{\alpha^2 + 3 \beta^2}{2 (\alpha^2 - \beta^2)^2} 
\left(1 + m \sqrt{\alpha^2 - \beta^2} \ \frac{K_0}{K_1} \right).
\end{eqnarray}
Let us also consider the energy
\begin{eqnarray}
\!\!\!\!\!\!\!\!\!\langle E \rangle&=&
\frac{1}{2 \pi} \int dp \ |\Phi(p)|^2 \sqrt{p^2 + m^2} \nonumber \\
&=&\frac{|A|^2 m^2}{2 \pi} \int dz \ \cosh^2 z
\exp\left(- 2 \alpha m \cosh z + 2 \beta m p \sinh z\right) \nonumber \\
&=&\frac{|A|^2}{4 \pi} \p_\alpha^2 K_0 = 
- \frac{\p_\alpha^2 K_0}{2 \p_\alpha K_0} =
\frac{\alpha}{\alpha^2 - \beta^2} 
\left(1 + m \sqrt{\alpha^2 - \beta^2} \ \frac{K_0}{K_1} \right) - 
\frac{1}{2 \alpha},
\end{eqnarray}
as well as the energy squared
\begin{eqnarray}
\langle E^2 \rangle = \langle p^2 \rangle + m^2 = 
\frac{m^2 \alpha^2}{\alpha^2 - \beta^2} +
\frac{\alpha^2 + 3 \beta^2}{2 (\alpha^2 - \beta^2)^2} 
\left(1 + m \sqrt{\alpha^2 - \beta^2} \ \frac{K_0}{K_1} \right).
\end{eqnarray}
Finally, using the same methods one can show that
\begin{equation}
\frac{\alpha}{2} \left\langle \frac{m^2}{E^3} \right\rangle + \alpha^2
\left\langle \frac{m^2}{p^2 + m^2} \right\rangle = \alpha^2 - \beta^2, 
\end{equation}
which then leads to
\begin{equation}
\langle x^2 \rangle = 
\frac{\alpha}{2} \left\langle \frac{m^2}{E^3} \right\rangle =
\alpha^2 - \beta^2 - 
\frac{2 \alpha \sqrt{\alpha^2 - \beta^2}}{K_1} 
\int_\alpha^\infty d\alpha' \ K_0\left(2 m \sqrt{{\alpha'}^2 - \beta^2}\right), 
\end{equation}

\section{Relation between Relativistic Quantum \\ Mechanics and Quantum Field 
Theory}

In this appendix we review the relation between relativistic quantum
mechanics and quantum field theory in the context of a simple free scalar field
theory in $(1+1)$ dimensions.

\subsection{Canonical Quantization}

Let us consider a free field theory for a real-valued massive scalar field 
$\varphi(x,t) \in \R$ with the Lagrangian
\begin{equation}
{\cal L} = \frac{1}{2} \left[\left(\p_t \varphi\right)^2 -  
\left(\p_x \varphi\right)^2 - m^2 \varphi^2\right].
\end{equation}
The momentum conjugate to the field $\varphi$ is given by
\begin{equation}
\Pi(x) = \frac{\delta {\cal L}}{\delta \p_t \varphi(x)} = \p_t \varphi(x),
\end{equation}
and the classical Hamilton density thus takes the form
\begin{equation}
{\cal H} = \Pi \p_t \varphi - {\cal L} = 
\frac{1}{2} \left[\Pi^2 + \left(\p_x \varphi\right)^2 + m^2 \varphi^2\right].
\end{equation}
Upon canonical quantization, the classical fields $\varphi(x)$ and $\Pi(x)$ are
replaced by field operators with the commutation relations
\begin{equation}
[\varphi(x),\varphi(x')] = [\Pi(x),\Pi(x')] = 0, \quad 
[\varphi(x),\Pi(x')] = i \delta(x - x').
\end{equation}
The classical Hamilton density then turns into the Hamilton operator
\begin{equation}
H = \int dx \ {\cal H} = \int dx \ 
\frac{1}{2} \left[\Pi^2 + \left(\p_x \varphi\right)^2 + m^2 \varphi^2\right].
\end{equation}

\subsection{Particle Spectrum}

In order to diagonalize the Hamiltonian, we go to momentum space by writing
\begin{equation}
\varphi(x) = \frac{1}{2 \pi} \int dp \ \widetilde\varphi(p) \exp(i p x), \quad
\Pi(x) = \frac{1}{2 \pi} \int dp \ \widetilde\Pi(p) \exp(i p x).
\end{equation}
The field operators in momentum space satisfy
\begin{equation}
\widetilde\varphi(p)^\dagger = \widetilde\varphi(-p), \quad 
\widetilde\Pi(p)^\dagger = \widetilde\Pi(-p),
\end{equation}
and obey the commutation relations
\begin{equation}
[\widetilde\varphi(p),\widetilde\varphi(p')] = 
[\widetilde\Pi(p),\widetilde\Pi(p')] = 0, \quad 
[\widetilde\varphi(p),\widetilde\Pi(p')] = 2 \pi i \delta(p + p').
\end{equation}
The Hamilton operator is then given by
\begin{equation}
H = \frac{1}{2 \pi} \int dp \ 
\frac{1}{2} \left[\widetilde\Pi^\dagger \widetilde\Pi + 
(p^2 + m^2) \widetilde\varphi^\dagger \widetilde\varphi\right].
\end{equation}

Let us now introduce particle creation and annihilation operators
\begin{eqnarray}
a(p)&=&\frac{1}{\sqrt{2}}\left[(p^2 + m^2)^{1/4} \widetilde\varphi(p) +
i (p^2 + m^2)^{-1/4} \widetilde\Pi(p)\right], \nonumber \\
a(p)^\dagger&=&\frac{1}{\sqrt{2}}\left[(p^2 + m^2)^{1/4} 
\widetilde\varphi(p)^\dagger - 
i (p^2 + m^2)^{-1/4} \widetilde\Pi(p)^\dagger\right],
\end{eqnarray}
which obey the commutation relations
\begin{equation}
[a(p),a(p')] = [a(p)^\dagger,a(p')^\dagger] = 0, \quad 
[a(p),a(p')^\dagger] = 2 \pi \delta(p - p'),
\end{equation}
The Hamilton operator then takes the form
\begin{equation}
H = \frac{1}{2 \pi} \int dp \ \sqrt{p^2 + m^2} 
\left[a(p)^\dagger a(p) + \pi \delta(0)\right],
\end{equation}
where the last term represents the divergent vacuum energy. The
vacuum state $|0\rangle$ is characterized by $a(p) |0\rangle = 0$ for all values
of $p$. The single particle states with momentum $p$ and energy 
$E(p) = \sqrt{p^2 + m^2}$ are given by
\begin{equation}
|p\rangle = a(p)^\dagger |0\rangle.
\end{equation}

\subsection{Localization of Particle States}

According to the standard rules of quantum mechanics, one may construct a
single particle position eigenstate
\begin{equation}
|x\rangle = \frac{1}{2 \pi} \int dp \ \exp(- i p x) |p\rangle.
\end{equation}
Introducing
\begin{equation}
a(x)^\dagger = \frac{1}{2 \pi} \int dp \ a(p)^\dagger \exp(- i p x),
\end{equation}
one then obtains
\begin{equation}
|x\rangle = a(x)^\dagger |0\rangle.
\end{equation}
According to our considerations in relativistic quantum mechanics, a wave packet
composed of such states moves and spreads in a manner that seems to violate
causality. Exactly the same behavior also arises in quantum field theory. 
Indeed, when we form single particle wave packets in the scalar field theory, 
they behave exactly as the ones in relativistic quantum mechanics that were
considered in section 4. 

In contrast to relativistic quantum mechanics, relativistic quantum field 
theory is based on the principle of locality. Hence, by construction, causality
cannot be violated. Indeed, as we will now see, the apparent violation of 
causality observed in wave packet spreading is due to the fact that single 
particle states cannot be localized in a finite region. This is a consequence of
the Reeh-Schlieder theorem \cite{Ree61}. The issues of particle localization 
have already been discussed by Newton and Wigner in 1949 \cite{New49}, have 
been investigated further, for example, in \cite{Rui81}, and continue to be a 
subject of controversial discussions \cite{Fle00,Hal01}. We notice that the 
operator
\begin{equation}
a(x)^\dagger = \frac{1}{2 \pi} \int dp \
\frac{1}{\sqrt{2}}\left[(p^2 + m^2)^{1/4} \widetilde\varphi(p)^\dagger - 
i (p^2 + m^2)^{-1/4} \widetilde\Pi(p)^\dagger\right] \exp(-i p x),
\end{equation}
is not localized at $x$, but is instead non-local. In fact, the single particle
position eigenstate $|x\rangle$ cannot be created from the vacuum by an
application of the field operators $\varphi(x)$ and $\Pi(x)$ and their
derivatives at the point $x$. Consequently, in quantum field theory single 
particles are inherently non-local objects. In non-relativistic quantum 
mechanics such subtleties do not arise and one interprets the state $|x\rangle$
as describing a single particle completely localized at the point $x$. When we
do the same in relativistic quantum mechanics, we encounter an apparent 
violation of causality 
\cite{Fle65,Heg74,Heg80,Heg85,Ros87,Mos90,Heg01,Bar03,Bus05,Bus06}. As the
discussion of field theory shows, causality is not really violated, since the 
particle itself is a non-local object. This should be kept in mind when one 
interprets the results obtained in the framework of relativistic quantum 
mechanics. Consequently, the state $|x\rangle$ should not be viewed as 
describing a particle localized at the point $x$, but should be associated with
the corresponding state in quantum field theory which is not localized.

\end{appendix}


\newpage

\begin{thebibliography}{10}

\bibitem{Gar95}
B.\ M.\ Garraway and K.-A.\ Suominen, Rep.\ Prog.\ Phys.\ 58 (1995) 365.

\bibitem{And08}
M.\ Andrews, Am.\ J.\ Phys.\ 76 (2008) 1102.

\bibitem{Ehr27}
P.\ Ehrenfest, Z.\ Phys.\ 45 (1927) 455.

\bibitem{Ree61}
H.\ Reeh and S.\ Schlieder, Nuovo Cim.\ 22 (1961) 1051.

\bibitem{Fle65}
G.\ N.\ Fleming, Phys.\ Rev.\ 139 (1965) 963. 

\bibitem{Heg74}
G.\ C.\ Hegerfeldt, Phys.\ Rev.\ D10 (1974) 3320.

\bibitem{Heg80}
G.\ C.\ Hegerfeldt and S.\ N.\ M.\ Ruijsenaars, Phys.\ Rev.\ D22 (1980) 377.

\bibitem{Heg85}
G.\ C.\ Hegerfeldt, Phys.\ Rev.\ Lett.\ 54 (1985) 2395.

\bibitem{Ros87}
B.\ Rosenstein and M.\ Usher, Phys.\ Rev.\ D36 (1987) 2381.

\bibitem{Mos90}
S.\ N.\ Mosley and J.\ E.\ G.\ Farina, J.\ Phys.\ A: Math.\ Gen.\ 23 (1990) 
3991.

\bibitem{Heg01}
G.\ C.\ Hegerfeldt, in ``Extensions of Quantum Theory'', edited by A.\ Horzela
and E.\ Kapuscik, published by Apeiron, Montreal (2001) 9.

\bibitem{Bar03}
N.\ Barat and J.\ C.\ Kimball, Phys.\ Lett.\ A308 (2003) 110.

\bibitem{Bus05}
F.\ Buscemi and G.\ Compagno, Phys.\ Lett.\ A334 (2005) 357.

\bibitem{Bus06}
F.\ Buscemi and G.\ Compagno, J.\ Phys.\ B: At.\ Mol.\ Opt.\ Phys.\ 39 (2006) 
695.

\bibitem{Leu65}
H.\ Leutwyler, Nuovo Cim.\ 37 (1965) 543.

\bibitem{Cur63}
D.\ G.\ Currie, T.\ F.\ Jordan, and E.\ C.\ G.\ Sudarshan, Rev.\ Mod.\ Phys.\ 
35 (1963) 350.

\bibitem{Rui80}
S.\ N.\ M.\ Ruijsenaars, Ann.\ Phys.\ 126 (1980) 399.

\bibitem{Rui86}
S.\ N.\ M.\ Ruijsenaars, Commun.\ Math.\ Phys.\ 110 (1987) 191.

\bibitem{Rui01}
S.\ N.\ M.\ Ruijsenaars, in ``Integrable Structures of Exactly Solvable
Two-Dimensional Models of Quantum Field Theory'', edited by
S.\ Pakuliak and G.\ von Gehlen, published by Kluwer Academic Publishers,
Netherlands (2001) 273.

\bibitem{Giu91}
C.\ Giunti, C.\ W.\ Kim, and U.\ W.\ Lee, Phys.\ Rev.\ D44 (1991) 3635.

\bibitem{Giu04}
C.\ Giunti, Foundations of Physics Letters 17 (2004) 103.

\bibitem{Bai72}
L.\ C.\ Baird, Am.\ J.\ Phys.\ 40 (1972) 327.

\bibitem{Bak73}
F.\ Bakke and H.\ Wergeland, Physica 69 (1973) 5.

\bibitem{Alm84}
C.\ Almeida and A.\ Jabs, Am.\ J.\ Phys.\ 52 (1984) 921.

\bibitem{Str06}
F.\ W.\ Strauch, Phys.\ Rev.\ A73 (2006) 069908.

\bibitem{Sch89}
G.\ Scharf, Finite Quantum Electrodynamics, Springer (1989).

\bibitem{New49}
T.\ D.\ Newton and E.\ P.\ Wigner, Rev.\ Mod.\ Phys.\ 21 (1949) 400.

\bibitem{Rui81}
S.\ N.\ M.\ Ruijsenaars, Ann.\ Phys.\ 137 (1981) 33.

\bibitem{Fle00}
G.\ N.\ Fleming, Philosophy of Science 67 (2000) 495.

\bibitem{Hal01}
H.\ Halverson, Philosophy of Science 68 (2001) 111.

\end{thebibliography}
\end{document}